\documentclass[journal]{IEEEtran}
\usepackage[cmex10]{amsmath}
\usepackage{graphicx,amssymb,amsthm,comment,bm,cite,url}
\usepackage{color}
\usepackage{boldline}
\usepackage{algorithm,algorithmic}
\usepackage{multirow}
\usepackage{cite}
\allowdisplaybreaks

\newcommand{\refeq}[1]{Eq. (\ref{eq:#1})}
\newcommand{\refeqs}[2]{Eqs. (\ref{eq:#1}) and (\ref{eq:#2})}

\newcommand{\refsec}[1]{Section \ref{sec:#1}}
\newcommand{\refsubsec}[1]{Subsection \ref{subsec:#1}}

\newcommand{\reffig}[1]{Fig. \ref{fig:#1}}

\newcommand{\reftab}[1]{Tab. \ref{tab:#1}}

\newcommand{\refalgo}[1]{Algorithm \ref{algo:#1}}

\def\Vec#1{\boldsymbol{\mathbf{#1}}}
\def\Argmax{\mathop{\rm argmax}}

\def\thline{\noalign{\hrule height 1.2pt}}

\ifCLASSINFOpdf
\else
\fi

\begin{document}
%
\title{
Many-to-Many Voice Transformer Network 
}
%
%
%

\author{Hirokazu~Kameoka,
Wen-Chin Huang,
Kou Tanaka,\\
Takuhiro Kaneko,
Nobukatsu Hojo,
and
Tomoki Toda
\thanks{H. Kameoka, K. Tanaka, 
T. Kaneko and N. Hojo are with NTT Communication Science Laboratories, Nippon Telegraph and Telephone Corporation, Atsugi, Kanagawa, 243-0198 Japan (e-mail: hirokazu.kameoka.uh@hco.ntt.co.jp) 
and W.-C. Huang and T. Toda are with Nagoya University.
}
\thanks{
This work was supported by JSPS KAKENHI 17H01763 and JST CREST Grant Number JPMJCR19A3, Japan.
}}

%
%

\markboth{
}%
{Shell \MakeLowercase{\textit{et al.}}: Bare Demo of IEEEtran.cls for Journals}
%



\maketitle

\begin{abstract}
This paper proposes a voice conversion (VC) method based on a sequence-to-sequence (S2S) learning framework, which enables simultaneous conversion of the voice characteristics, pitch contour, and duration of input speech. We previously proposed an S2S-based VC method using a transformer network architecture called the voice transformer network (VTN). The original VTN was designed to learn only a mapping of speech feature sequences from one speaker to another. The main idea we propose is an extension of the original VTN that can simultaneously learn mappings among multiple speakers. This extension called the many-to-many VTN makes it able to fully use available training data collected from multiple speakers by capturing common latent features that can be shared across different speakers. It also allows us to introduce a training loss called the identity mapping loss to ensure that the input feature sequence will remain unchanged when the source and target speaker indices are the same. Using this particular loss for model training has been found to be extremely effective in improving the performance of the model at test time. We conducted speaker identity conversion experiments and found that our model obtained higher sound quality and speaker similarity than baseline methods. We also found that our model, with a slight modification to its architecture, could handle any-to-many conversion tasks reasonably well.
\end{abstract}


%
\IEEEpeerreviewmaketitle

\section{Introduction}
\label{sec:intro}

Voice conversion (VC) is a technique to modify
the voice characteristic and speaking style of an input utterance 
without changing the linguistic content.
Potential applications of VC include
speaker-identity modification \cite{Kain1998short}, 
speaking aids \cite{Kain2007short,Nakamura2012short}, 
speech enhancement \cite{Inanoglu2009short,Toda2012short}, 
and accent conversion \cite{Felps2009short}.

Many conventional VC methods 
use parallel utterances of source and target speech to 
train acoustic models for feature mapping. 
The training process typically
consists of computing acoustic features from source and target utterances,
using dynamic time warping (DTW) to align parallel utterances,
and training the acoustic model 
that describes the mappings between 
the source features and the target features. 
Examples of acoustic models include Gaussian mixture models (GMMs) \cite{Stylianou1998short,Toda2007short}, partial least square regression \cite{Helander2010}, 
dynamic kernel partial least square regression \cite{Helander2011}, 
frequency warping \cite{Tian2017},
non-negative matrix factorization \cite{Takashima2012},
group sparse representation \cite{Sisman2017},
fully-connected deep neural networks (DNNs) \cite{Desai2010short,Mohammadi2014short},
and long-short term memory networks (LSTMs) \cite{Sun2015,Ming2016lstm},
just to name a few. 
Recently, attempts have also been made to 
formulate 
non-parallel methods using deep generative models.
These include methods based on
variational autoencoders \cite{Hsu2016short, Hsu2017short, YSaito2018b, Kameoka2019IEEETransshort_ACVAE-VC, Tobing2019}, generative adversarial networks \cite{Kaneko2018, Kameoka2018SLT_StarGAN-VC}, and
flow-based models \cite{Serra2019}.
Interested readers are referred to an excellent review article \cite{Sisman2020} on the recent trends in VC research.

Although most of the methods 
mentioned above are successful in converting the local spectral features, 
they have difficulty in converting suprasegmental features that reflect long-term dependencies,  
such as the fundamental frequency ($F_0$) contour, duration and rhythm of the input speech.
This is because acoustic models in these methods
can only describe
mappings between local features.
However, since these suprasegmental features 
are as important factors as local features
that characterize speaker identities and speaking styles, 
it would be desirable if these features could also be converted more flexibly.
Although some attempts have been made to 
achieve conversions of $F_0$ and energy contours through 
their continuous wavelet transform representations \cite{Sanchez2014,Ming2016exemplar,Ming2016lstm,Luo2017b,Sisman2019}, 
they do not address conversions of duration and rhythm.
One solution to overcome this 
limitation
would be to adopt a sequence-to-sequence (seq2seq or S2S) model,
as it has a powerful ability to 
learn mappings between sequential data of variable lengths
by capturing and using long-range dependencies. 

S2S models
\cite{Sutskever2014short,Chorowski2015NIPS}
have recently been applied 
with notable success 
in many tasks 
such as machine translation, automatic speech recognition (ASR) 
\cite{Chorowski2015NIPS} and
text-to-speech (TTS) \cite{Wang2017short,Arik2017ashort,Arik2017bshort,Sotelo2017short,Tachibana2018short,Ping2018ICLRshort}.
They are
composed mainly of two elements: an encoder and decoder.
The encoder encodes an input sequence to 
its latent representation in the form of hidden state vectors
whereas the decoder generates an output sequence according to 
this latent representation. 
With the original S2S model, 
all input sequences are encoded into 
a single context vector of a fixed dimension.
One problem with this is that 
the ability of the model to capture long-range dependencies
can be limited 
especially when input sequences are long.
To overcome this limitation,
a mechanism called attention \cite{Luong2015short} has been introduced, 
which allows the network to learn where to pay attention in the input sequence
when producing each item in the output sequence.

While 
recurrent neural networks (RNNs)
have initially been used as the default option for
designing the encoder and decoder networks in S2S models, 
recent work has shown that
convolutional neural network (CNN)-based architectures 
also have excellent potential
for capturing long-term dependencies \cite{Gehring2017arXivshort}.
Subsequently, 
yet another type of architecture
called the transformer has been proposed \cite{Vaswani2017NIPSshort}, which
uses neither convolution nor recurrent layers in its network, 
but only the mechanism of attention. 
In particular, it uses multi-head self-attention layers
to design the two networks. 
Self-attention is a type of attention mechanism, 
which offers an efficient way to relate different positions of a given sequence. 
The multi-head self-attention mechanism 
splits each vector in a sequence into smaller parts 
and then computes the self-attention over the sequence of each part in parallel.
Unlike RNNs,
both these architectures have the advantage that
they are suitable for parallel computations using GPUs. 

Several VC methods based on S2S models have already been proposed \cite{Miyoshi2017short,Zhang2018short,Zhang2019short,Biadsy2019short}.
We also previously proposed VC methods based on 
S2S models with 
RNN \cite{Tanaka2019ICASSP_AttS2S-VC}, CNN \cite{Kameoka2018IEEE-TASLP_ConvS2S-VC} and transformer architectures \cite{Huang2020Interspeech_VTN}. 
In our most recent work \cite{Huang2020Interspeech_VTN}, 
we proposed 
a VC method based on a transformer architecture, which we  
called the voice transformer network (VTN).
Through this work, it transpired that 
the model trained from scratch
did not perform as expected 
when the amount of training data was limited.
To address this, 
we introduced a TTS pretraining technique 
to provide a good initialization for fast and sample-efficient VC model 
training with the aid of text-speech pair data, thus reducing 
the parallel data size requirement 
and training time.

In this paper, we propose several ideas to make the VTN perform well 
even without TTS pretraining.
One limitation with regular S2S models including the VTN 
is that they can only learn a mapping from one speaker to another.
When parallel utterances of multiple speakers are available, 
naively preparing and training a different model  
independently
for each speaker pair
would be inefficient, 
since
the model for a particular speaker pair 
fails to use the training data of the other speakers for its training.
To fully utilize available training data in multiple speakers, 
the main idea we propose is 
extending the VTN so that it can simultaneously learn 
mappings among multiple speakers.
We call this extended version the many-to-many VTN.
The idea of this many-to-many extension
was triggered by 
our previous studies on non-parallel VC methods \cite{Kameoka2020arXiv_AStarGAN-VC} that
showed the significant superiority of a many-to-many VC method
\cite{Kameoka2018SLT_StarGAN-VC}
over its one-to-one counterpart
\cite{Kaneko2018}.
This suggests the possibility 
that 
in learning mappings between a certain speaker pair
the data of other speakers may also be useful. 
However, the effect this extension would have on parallel VC tasks based on S2S models was nontrivial.
In this regard, we consider the present contribution important.
We further propose
several ideas, 
including 
the identity mapping loss,
attention windowing, and
settings needed to achieve any-to-many and real-time conversions. 
We show that the performance improvement achieved by the many-to-many extension owes not only to the fact that it can leverage more training data than the one-to-one counterpart but also to the identity mapping loss, which is only available in the many-to-many version.
We also show that the pre-layer normalization (Pre-LN) architecture, 
proposed in \cite{Wang2019ACL,Xiong2020ICML},
is effective in both the original (one-to-one) and many-to-many VTNs.

\section{Related work}

Several VC methods based on S2S models have already been proposed,
including the ones we proposed previously. 
Regular S2S models usually require large-scale parallel corpora for training. 
However, collecting parallel utterances is often costly and non-scalable.
Therefore, in VC tasks using S2S models, one challenge is how to make them work with a limited amount of training data.

One solution is to use text labels as auxiliary information for model training. 
Miyoshi et al. proposed a VC model combining an S2S model and
acoustic models for ASR and TTS \cite{Miyoshi2017short}. 
Zhang et al. proposed an S2S-based VC model guided by an ASR system \cite{Zhang2018short}.
Subsequently, Zhang et al. proposed a shared S2S model for
TTS and VC tasks \cite{Zhang2019short}.
Recently, Biadsy et al. proposed an end-to-end VC model called ``Parrotron'', which uses a multitask learning strategy to train the encoder and decoder along with an ASR model \cite{Biadsy2019short}.
Our VTN \cite{Huang2020Interspeech_VTN} is another example, 
which relies on TTS pretraining using text-speech pair data.  

The proposed many-to-many VTN differs from the above methods
in that it does not rely on ASR or TTS models and requires no text annotations for model training.

\section{Voice Transformer Network}
\label{sec:VTN}

\subsection{Feature extraction and normalization}
\label{subsec:feature}

Following our previous work \cite{Kameoka2018IEEE-TASLP_ConvS2S-VC},
in this work, we choose to use the mel-cepstral coefficients (MCCs) \cite{Fukada1992short}, log $F_0$, aperiodicity, and voiced/unvoiced indicator of speech as the acoustic features to be converted.
Once acoustic features have successfully been converted,
we can use either a conventional vocoder or one of 
the recently developed high-quality neural vocoders \cite{vandenOord2016short,Tamamori2017short,Kalchbrenner2018short,Mehri2016short,Jin2018short,vandenOord2017short,Ping2019short,Prenger2018short,Kim2018short,Wang2018short,Tanaka2018short,Kumar2019,Yamamoto2020}
to generate the signals of converted speech. 

To obtain an acoustic feature vector, 
we follow the same procedure in \cite{Kameoka2018IEEE-TASLP_ConvS2S-VC}.
Namely, we first use WORLD \cite{Morise2016short} to analyze 
the spectral envelope, log $F_0$,  
coded aperiodicity, and 
voiced/unvoiced indicator 
within each time frame of a speech utterance, 
then compute $I$
MCCs from the extracted spectral envelope, 
and finally construct an acoustic feature vector 
by stacking
the MCCs,
log $F_0$, 
coded aperiodicity, 
and 
voiced/unvoiced indicator.
Each acoustic feature vector thus consists of $I+3$ elements.
At training time, 
we normalize 
each element $x_{i,n}$ $(i=1,\ldots,I)$ of the MCCs
and log $F_0$ $x_{I+1,n}$ at frame $n$ 
to
$x_{i,n} \leftarrow (x_{i,n} - \mu_i)/\sigma_i$
where $i$, $\mu_i$, and $\sigma_i$
denote the feature index,
mean, and standard deviation of the $i$-th feature 
within all the voiced segments of the training samples of the same speaker.

As with our previous work \cite{Kameoka2018IEEE-TASLP_ConvS2S-VC},
we found it useful to use a similar trick introduced in 
previous work \cite{Wang2017} to accelerate and stabilize training and inference.
Namely, we divide the acoustic feature sequence obtained above
into non-overlapping segments of equal length $r$
and use the stack of the acoustic feature vectors 
in each segment as a new feature vector
so that 
the new feature sequence becomes $r$ times shorter than the original feature sequence. 
\begin{figure*}[t!]
\centering
\begin{minipage}[t]{.3\linewidth}
  \centerline{\includegraphics[height=8.208cm]{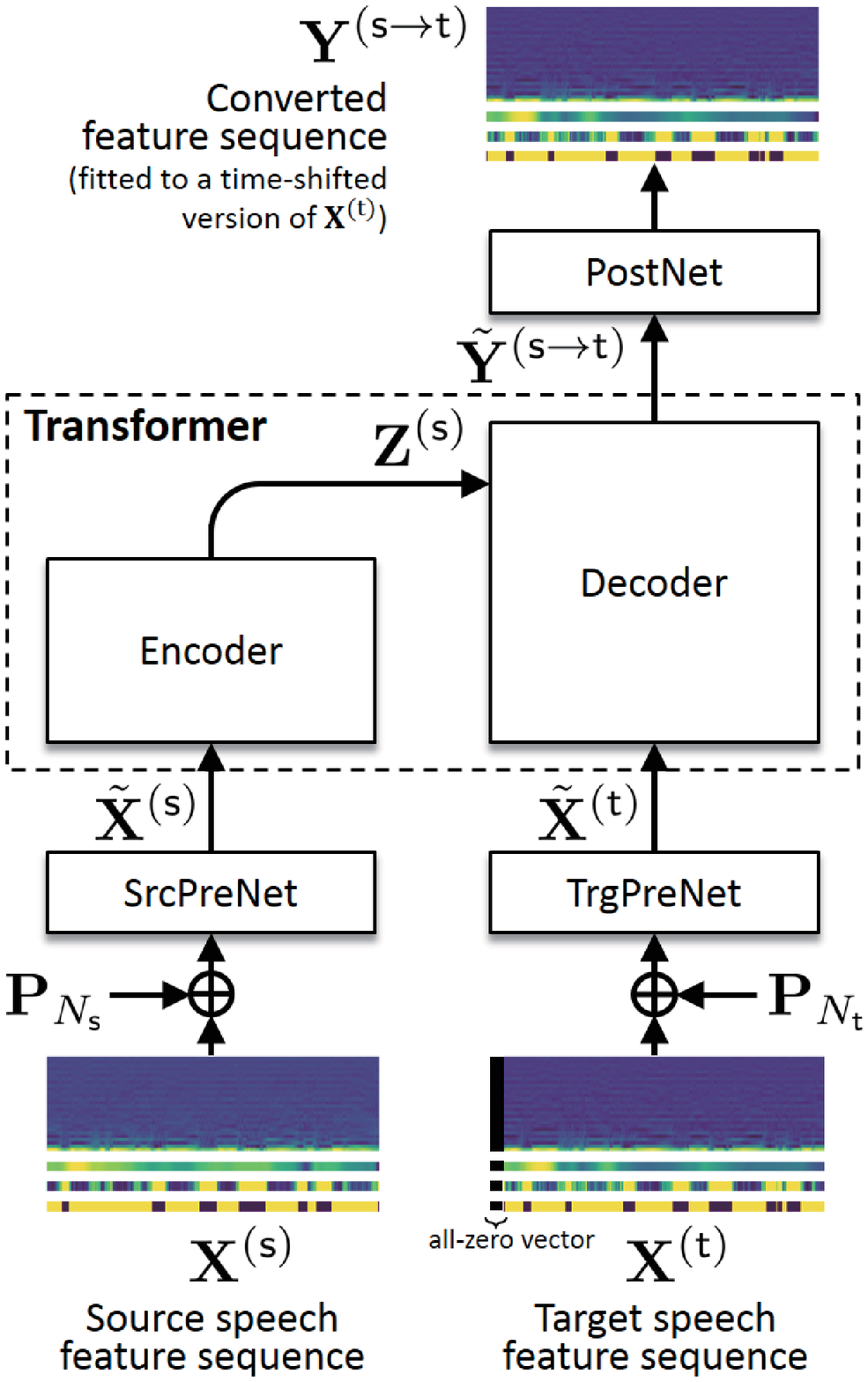}}
  \vspace{-1ex}
  \caption{Overall structure of VTN}
  \label{fig:VTN}
\end{minipage}
\hspace{2ex}
\begin{minipage}[t]{.25\linewidth}
  \centerline{\includegraphics[height=5.0676cm]{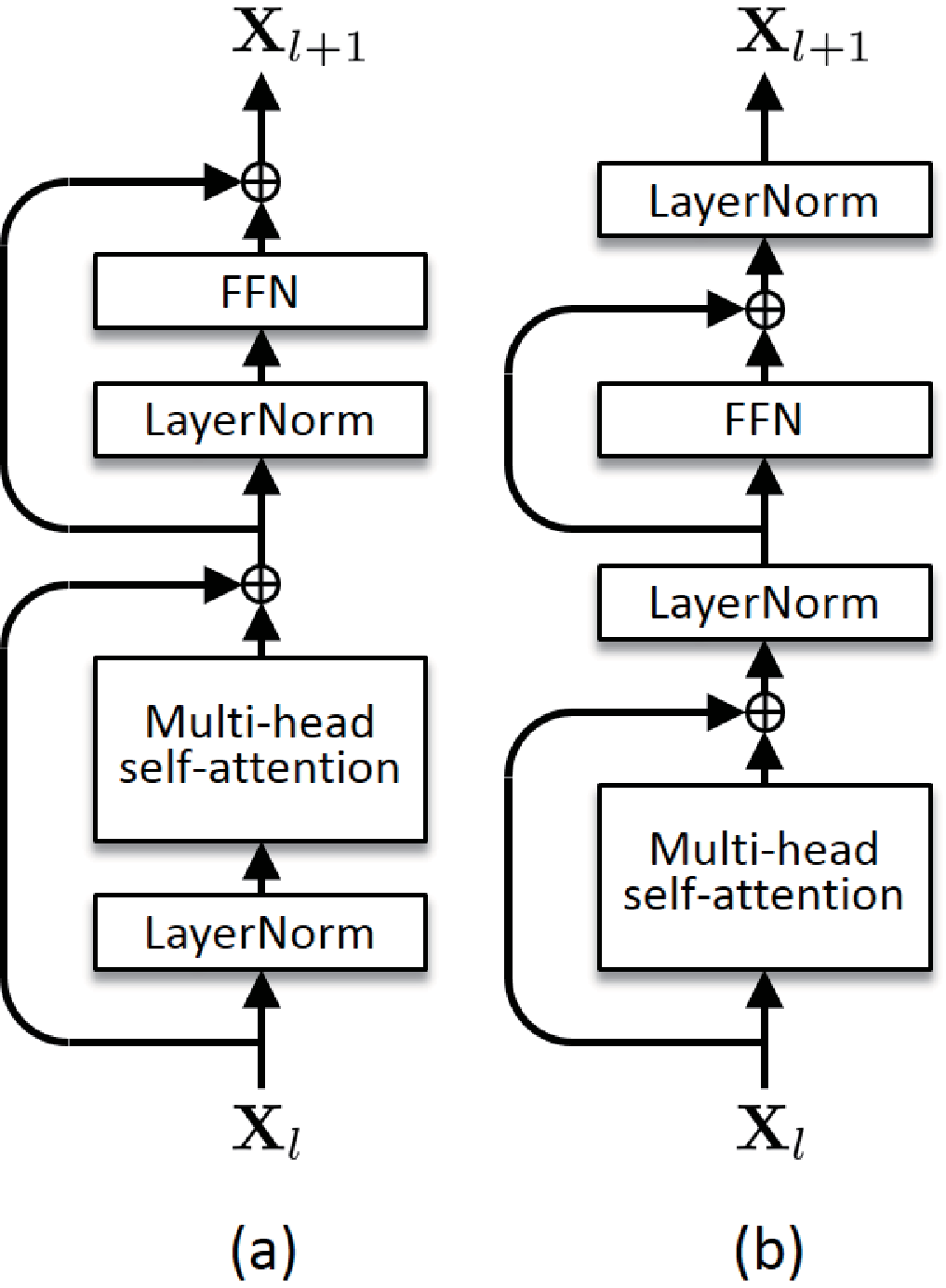}}
  \vspace{-1ex}
  \caption{Encoder layers with (a) Pre-LN and (b) Post-LN architectures}
  \label{fig:encoder}
\end{minipage}
\hspace{2ex}
\begin{minipage}[t]{.3\linewidth}
  \centerline{\includegraphics[height=7.32cm]{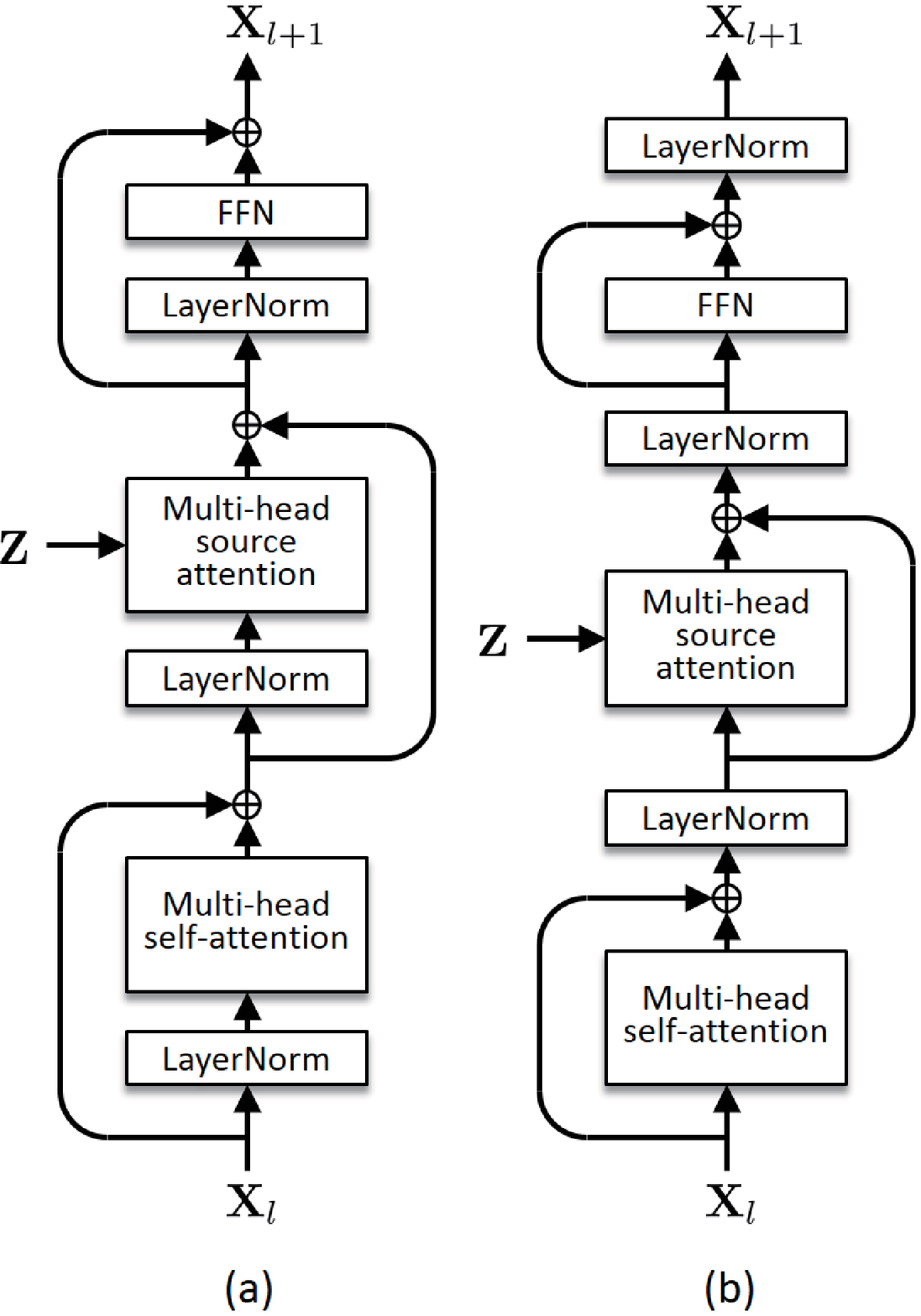}}
  \vspace{-1ex}
  \caption{Decoder layers with (a) Pre-LN and (b) Post-LN architectures}
  \label{fig:decoder}
\end{minipage}
\end{figure*}

\subsection{Model}
\label{subsec:model}
We hereafter use
$\Vec{X}^{(\mathsf{s})}=[\Vec{x}_1^{(\mathsf{s})},\ldots,
\Vec{x}_{N_{\mathsf s}}^{(\mathsf{s})}]\in \mathbb{R}^{D\times N_{\mathsf s}}$ and 
$\Vec{X}^{(\mathsf{t})}=[\Vec{x}_1^{(\mathsf{t})},\ldots,
\Vec{x}_{N_{\mathsf t}}^{(\mathsf{t})}]\in \mathbb{R}^{D\times N_{\mathsf t}}$
to denote the source and target speech feature sequences of non-aligned parallel utterances,
where $N_{\mathsf s}$ and $N_{\mathsf t}$ denote the lengths of the two sequences and 
$D$ denotes the feature dimension. 
The
VTN \cite{Huang2020Interspeech_VTN} 
has an encoder-decoder structure with
a transformer architecture \cite{Vaswani2017NIPSshort}
that maps $\Vec{X}^{(\mathsf{s})}$ to $\Vec{X}^{(\mathsf{t})}$ (\reffig{VTN}).
The encoder is expected to extract 
contextual information from source speech,
and the decoder 
produces the target speech feature sequence
according to the contextual information the encoder has generated.
Unlike RNN- and CNN-based S2S models,
the transformer model 
does not have any sense of the order of the elements in a sequence.
Thus, 
the sinusoidal position encodings \cite{Vaswani2017NIPSshort}, 
denoted by
$\Vec{P}_{N_{\mathsf s}}\in\mathbb{R}^{D\times N_{\mathsf s}}$ and 
$\Vec{P}_{N_{\mathsf t}}\in\mathbb{R}^{D\times N_{\mathsf t}}$, 
are first added to the source and target feature sequences
to make the model ``aware'' of 
the position at which an each element in the sequences is located. 
The source and target feature sequences
are then passed through convolutional prenets, which we call the source and target prenets, 
before being fed into the encoder and decoder.
The output $\tilde{\Vec{Y}}{}^{(\mathsf{s}\rightarrow\mathsf{t})}$
from the decoder is finally passed through a 
convolutional postnet before producing the final output 
$\Vec{Y}^{(\mathsf{s}\rightarrow\mathsf{t})}$.
The two prenets and the postnet,
each consisting of three convolution layers,
are used to 
capture and express the local dynamics in 
source and target speech and
convert input sequences into sequences of the same lengths.
In the following, we use 
$\tilde{\Vec{X}}{}^{(\mathsf{s})}\in\mathbb{R}^{d\times N_{\mathsf s}}$ 
and 
$\tilde{\Vec{X}}{}^{(\mathsf{t})}\in\mathbb{R}^{d\times N_{\mathsf t}}$ 
to denote the outputs from the source and target prenets, respectively,
where $d$ is the output channel number of each prenet.

\subsubsection{Encoder}

The encoder takes $\tilde{\Vec{X}}{}^{(\mathsf{s})}$ as the input and
produces a context vector sequence 
$\Vec{Z}^{(\mathsf{s})}=
[\Vec{z}_1^{(\mathsf{s})},\ldots,
\Vec{z}_{N_{\mathsf s}}^{(\mathsf{s})}
]
\in\mathbb{R}^{d\times N_{\mathsf s}}$.
The encoder consists of $L$ identical layers, each of which has 
self-attention (SA) and position-wise fully-connected feed forward network (FFN) sub-layers.
Residual connections and layer normalizations are applied in addition to the two sub-layers.

\noindent
{\bf Multi-head self-attention sub-layer:}
By using $\Vec{X}\in\mathbb{R}^{d\times N}$ and $\Vec{Y}\in\mathbb{R}^{d\times N}$
to denote the input and output sequences of length $N$ of an SA sub-layer,
the process $\Vec{Y}=\mathsf{SA}(\Vec{X})$,
by which $\Vec{Y}$ is produced, 
is given as
\begin{align}
[\Vec{Q};\Vec{K};\Vec{V}] &= \Vec{W}_1\Vec{X} \in\mathbb{R}^{3d\times N},\label{eq:mhsa_1}\\
&{\rm where}
\left\{
\begin{array}{l}
\Vec{Q} = [\Vec{Q}_1;\ldots;\Vec{Q}_H]\\ 
\Vec{K} = [\Vec{K}_1;\ldots;\Vec{K}_H]\\ 
\Vec{V} = [\Vec{V}_1;\ldots;\Vec{V}_H]\\ 
\end{array},
\right.
\label{eq:mhsa_2}\\
\Vec{A}_h &= \mathsf{softmax}
\big(\textstyle{\frac{\Vec{K}_h^{\mathsf T}\Vec{Q}_h}{\sqrt{d}}}\big)
~(h=1,\ldots,H),\label{eq:mhsa_5}\\
\Vec{Y} &= \Vec{W}_2 [\Vec{V}_1 \Vec{A}_1; \ldots; \Vec{V}_H \Vec{A}_H],
\label{eq:mhsa_6}
\end{align}
where 
$\Vec{W}_1\in\mathbb{R}^{3d \times d}$ and 
$\Vec{W}_2\in\mathbb{R}^{d\times d}$ 
are learnable weight matrices,
$\mathsf{softmax}$ denotes a softmax operation performed on the first axis,
$H$ denotes the number of heads, and
$[;]$ denotes vertical concatenation of matrices (or vectors) with compatible sizes.
Intuitively, this process can be understood as follows.
First, an input vector sequence is converted into three types of vector sequences 
with the same shape,
which can be metaphorically 
interpreted as the queries and the key-value pairs in a hash table.
Each of the three vector sequences is further split into 
$H$ homogeneous vector sequences with the same shape. 
By using the query and key pair, 
\refeq{mhsa_5} computes a self-attention matrix, whose element 
measures
how contextually similar each pair of vectors is in the given sequence $\Vec{X}$.
The splitting 
into $H$ heads
allows us to measure self-simiarity in terms of $H$ different types of context.
The $n$-th column of 
$\Vec{V}_h\Vec{A}_h$ in \refeq{mhsa_6}
can be seen as a new feature vector given by
activating
the value vectors at all the positions 
that are similar to the current position $n$ in terms of context $h$
and adding them together.
\refeq{mhsa_6} finally produces the output sequence $\Vec{Y}$ after
combining all these feature vector sequences using learnable weights.

\noindent
{\bf Position-wise feed forward network sub-layer:}
By using $\Vec{X}\in\mathbb{R}^{d\times N}$ 
and $\Vec{Y}\in\mathbb{R}^{d\times N}$
again to denote the input and output sequences of length $N$ of an FFN sub-layer,
the process $\Vec{Y} = \mathsf{FFN}(\Vec{X})$,
by which $\Vec{Y}$ is produced, 
is given as 
\begin{align}
\Vec{Y} = \Vec{W}_4 \phi(\Vec{W}_3\Vec{X} + \Vec{B}_3) + \Vec{B}_4,
\end{align}
where 
$\Vec{W}_3\in\mathbb{R}^{d'\times d}$, $\Vec{W}_4\in\mathbb{R}^{d\times d'}$ are 
learnable weight matrices,
$\Vec{B}_3=[\Vec{b}_3,\ldots,\Vec{b}_3] \in\mathbb{R}^{d'\times N}$ 
and 
$\Vec{B}_4=[\Vec{b}_4,\ldots,\Vec{b}_4]\in\mathbb{R}^{d\times N}$
are bias matrices, each consisting of 
identical learnable column vectors,
and $\phi$ denotes an elementwise nonlinear activation function
such as the rectified linear unit (ReLU) and
gated linear unit (GLU) functions.

\noindent
{\bf Layer normalization sub-layers:} 
Recent work has shown that 
the location of the layer normalization in the transformer architecture
affects the speed and stability of the training process 
as well as the performance of the trained model
\cite{Wang2019ACL,Xiong2020ICML}.
While the original transformer architecture places layer normalization 
after the SA and FFN sub-layers, 
the architectures presented
in the subsequent work
\cite{Wang2019ACL,Xiong2020ICML} 
are designed to 
place it before them,
as illustrated in \reffig{encoder}.
To distinguish between these two architectures,
we refer to the former and latter as
post-layer normalization (Post-LN)
and Pre-LN architectures, respectively.
We show later how differently these architectures 
actually performed in our experiments.
Note that when we say we apply layer normalization  
to an input vector sequence, say
$\Vec{X}=[\Vec{x}_1,\ldots,\Vec{x}_N]$, 
we mean applying layer normalization
to all the vectors $\Vec{x}_1,\ldots,\Vec{x}_N$
treated as mini-batch samples.

If we use $\Vec{X}_l$ and $\Vec{X}_{l+1}$
to denote the input and output of the $l$-th
encoder layer (with the PreLN architecture), 
the process $\Vec{X}_{l+1} = \mathsf{Enc}_l(\Vec{X}_l)$
of the $l$-the layer
is given by
\begin{align}
\Vec{U} &= \Vec{X}_{l} + \mathsf{SA}(\mathsf{LayerNorm}_1(\Vec{X}_l)),\\
\Vec{X}_{l+1} &= \Vec{U} + \mathsf{FFN}(\mathsf{LayerNorm}_2(\Vec{U})),
\end{align}
where $\mathsf{LayerNorm}_1$ and $\mathsf{LayerNorm}_2$
denote different LN sub-layers. 
As described above, 
each layer 
has learnable parameters in  
the SA and FFN sub-layers
and the two LN sub-layers.
The layer implemented as above
is particularly attractive 
in that 
it is able to relate all the positions in the entire input sequence
using only a single layer. 
This is in contrast to a regular convolution layer, which
is only able to relate local positions near each position.

\subsubsection{Decoder}

The decoder takes $\Vec{Z}^{(\mathsf{s})}$ and 
$\tilde{\Vec{X}}{}^{(\mathsf{t})}$ as the inputs and
produces a converted feature sequence 
$\tilde{\Vec{Y}}{}^{(\mathsf{s}\rightarrow\mathsf{t})}=
[\tilde{\Vec{y}}_1^{(\mathsf{s}\rightarrow\mathsf{t})},
\ldots,
\tilde{\Vec{y}}_{N_{\mathsf t}}^{(\mathsf{s}\rightarrow\mathsf{t})}]
\in
\mathbb{R}^{d\times N_{\mathsf t}}$.
Similar to the encoder,
the decoder consists of $L$ identical layers, each of which has 
SA and FFN sub-layers, residual connections, and 
layer normalization sub-layers. 
In addition to these sub-layers, each layer has 
a multi-head target-to-source attention (TSA) sub-layer, as illustrated in \reffig{decoder},
whose role is to find 
which position in the source feature sequence
contextually corresponds to 
each position in the target feature sequence
and convert 
the context vector sequence according to 
the predicted corresponding positions.

\noindent
{\bf Multi-head target-to-source attention sub-layer:}
By using $\Vec{X}\in\mathbb{R}^{d\times N}$ and 
$\Vec{Y}\in\mathbb{R}^{d\times N}$
to denote the input and output sequences of length $N$
of the TSA sub-layer,
the process $\Vec{Y}=\mathsf{TSA}(\Vec{X},\Vec{Z})$,
by which $\Vec{Y}$ is produced, 
is given 
in the same manner as the SA sub-layer
with the only difference being that 
the key and value pair $(\Vec{K},\Vec{V})$ is 
computed using the output $\Vec{Z}$ from the encoder:
\begin{align}
\Vec{Q} &= \Vec{W}_5\Vec{X},
\\
[\Vec{K};\Vec{V}] &= \Vec{W}_6\Vec{Z},
\label{eq:mhtsa_1}\\
&{\rm where}
\left\{
\begin{array}{l}
\Vec{Q}=[\Vec{Q}_1;\ldots;\Vec{Q}_H]
\\
\Vec{K}=[\Vec{K}_1;\ldots;\Vec{K}_H]
\\
\Vec{V}=[\Vec{V}_1;\ldots;\Vec{V}_H]
\end{array},
\right.\label{eq:mhtsa_2}\\
\Vec{A}_h &= \mathsf{softmax}
\big(\textstyle{\frac{\Vec{K}_h^{\mathsf T}\Vec{Q}_h}{\sqrt{d}}}\big)
~(h=1,\ldots,H),\label{eq:mhtsa_6}\\
\Vec{Y} &= \Vec{W}_7[\Vec{V}_1 \Vec{A}_1;\ldots;\Vec{V}_H \Vec{A}_H],
\label{eq:mhtsa_7}
\end{align}
where
$\Vec{W}_5\in\mathbb{R}^{d\times d}$, 
$\Vec{W}_6\in\mathbb{R}^{2d\times d}$, and 
$\Vec{W}_7\in\mathbb{R}^{d\times d}$
are learnable weight matrices.
Analogously to the SA sub-layer,
\refeq{mhtsa_6} is used to compute a TSA matrix
using the query and key pair,
where 
the $(n,m)$-th element indicates the similarity
between the $n$-th and $m$-th frames of source and target speech.
The peak trajectory of $\Vec{A}_h$ can thus be interpreted as a time warping
function that associates the frames of the source speech with those of the target speech.
The splitting into $H$ heads
allows us to measure the simiarity in terms of $H$ different types of context.
$\Vec{V}_h\Vec{A}_h$ in \refeq{mhtsa_7}
can be thought of as a time-warped version of $\Vec{V}_h$ in terms of context $h$.
\refeq{mhtsa_7} finally produces the output sequence $\Vec{Y}$ after
combining all these time-warped feature sequences using learnable weights.

All the other sub-layers are defined in the same way as the encoder.
The overall structures of the decoder layers with
the PreLN and PostLN architectures
are depicted in \reffig{decoder}.
If we use $\Vec{X}_{l}$ and $\Vec{X}_{l+1}$
to denote the input and output of
the $l$-th decoder layer (with the PreLN architecture), 
the process $\Vec{X}_{l+1} = \mathsf{Dec}(\Vec{X}_{l},\Vec{Z})$
of the $l$-th layer
is given by
\begin{align}
\Vec{U}_1 &= \Vec{X}_{l} + \mathsf{SA}(\mathsf{LayerNorm}_1(\Vec{X}_{l})),\\
\Vec{U}_2 &= \Vec{U}_1 + \mathsf{TSA}(\mathsf{LayerNorm}_2(\Vec{U}_1), \Vec{Z}),\\
\Vec{X}_{l+1} &= \Vec{U}_2 + \mathsf{FFN}(\mathsf{LayerNorm}_3(\Vec{U}_2)).
\end{align}
Note that each layer
has learnable parameters in  
the SA, FFN, and TSA sub-layers 
and the three LN sub-layers.

\subsubsection{Autoregressive structure}

Since the target feature sequence 
$\Vec{X}{}^{(\mathsf{t})}$ is of course not accessible at test time, 
we would want to use a feature vector that the decoder has generated 
as the input to the decoder for the next time step so that 
feature vectors can be generated one-by-one in a recursive manner.
To allow the model to behave in this manner, we must first take care 
that 
the decoder must not be allowed to use future information 
about the target feature vectors
when producing 
an output vector at each time step. 
This can be ensured by simply 
constraining the convolution layers in the target prenet to be causal
and 
replacing \refeq{mhsa_5} in all the SA sub-layers in the decoder
with
\begin{align}
\Vec{A}_h &= \mathsf{softmax}
\big(\textstyle{\frac{\Vec{K}_h^{\mathsf T}\Vec{Q}_h}{\sqrt{d}}}+\Vec{E}\big),
\label{eq:mhsa_5_causal}
\end{align}
where $\Vec{E}$ is a matrix whose $(n,n')$-th element 
is given by
\begin{align}
e_{n,n'}=
\begin{cases}
0&(n\le n')\\
-\infty&(n>n')
\end{cases},
\end{align}
so that the
predictions for position $n$ can depend only on the known outputs at positions less than $n$.
Second, an all-zero vector is appended to the left end of $\Vec{X}{}^{(\mathsf{t})}$ 
\begin{align}
\Vec{X}{}^{(\mathsf{t})}\leftarrow [\Vec{0}, \Vec{X}{}^{(\mathsf{t})}],
\end{align}
so that the recursion always begins with the all-zero vector.
Third, the output sequence $\Vec{Y}^{(\mathsf{s}\rightarrow\mathsf{t})}$ must correspond to 
a time-shifted version of $\Vec{X}{}^{(\mathsf{t})}$ so that at each time step the decoder 
will be able to predict the target speech feature vector that is likely to appear at the next time step.
To this end, we include an $L_1$ loss
\begin{align}
\mathcal{L}_{\mathsf{main}} = 
{\textstyle \frac{1}{M}}
\| [\Vec{Y}^{(\mathsf{s}\rightarrow\mathsf{t})}]_{:,1:M} - [\Vec{X}^{(\mathsf{t})}]{}_{:,2:M+1} \|_1,
\label{eq:decloss_pairwise}
\end{align}
in the training loss to be minimized, 
where 
we have used the colon operator $:$ to specify the range of indices
of the elements in a matrix we wish to extract.
For ease of notation, we use $:$ to represent 
all elements along an axis.

\subsection{Constraints on Attention Matrix}
\label{subsec:AL}

Since 
the alignment path between parallel utterances must lie close to the diagonal,
the diagonal region in the attention matrices in each TSA sub-layer in the decoder should be dominant.
By imposing such a restriction, the search space during training can be significantly reduced, 
thus significantly reducing the training effort.
One way to force the attention matrices 
to be diagonally dominant 
involves introducing a diagonal attention loss (DAL) \cite{Tachibana2018short}:
\begin{align}
\mathcal{L}_{\mathsf{dal}} = 
{\textstyle \frac{1}{NMLH} \sum_{l}\sum_{h}}
\|\Vec{G}_{N_{\mathsf s}\times N_{\mathsf t}}\odot \Vec{A}_{l,h}\|_1,
\end{align}
where 
$\Vec{A}_{l,h}$ denotes 
the target-to-source attention matrix of the $h$-th head in the TSA sub-layer 
in the $l$-th decoder layer,
$\odot$ denotes an elementwise product, and 
$\Vec{G}_{N_{\mathsf s}\times N_{\mathsf t}}
\in\mathbb{R}^{N_{\mathsf s}\times N_{\mathsf t}}$ 
is a non-negative weight matrix 
whose $(n,m)$-th element $w_{n,m}$ is defined as 
$w_{n,m} = 1- e^{-(n/N_{\mathsf s}-m/N_{\mathsf t})^2/2\nu^2}$.

\subsection{Training loss}
\label{subsec:training}

Given examples of parallel utterances, 
the total training loss for the VTN to be minimized is given as
\begin{align}
\mathcal{L} = 
\mathbb{E}_{\Vec{X}{}^{(\mathsf{s})},\Vec{X}{}^{(\mathsf{t})}}
\left\{
\mathcal{L}_{\mathsf{main}}
+
\lambda_{\mathsf{dal}}
\mathcal{L}_{\mathsf{dal}}
\right\},
\end{align}
where 
$\mathbb{E}_{\Vec{X}{}^{(\mathsf{s})},\Vec{X}{}^{(\mathsf{t})}}\{\cdot\}$ denotes 
the sample mean over all the training examples and 
$\lambda_{\mathsf{dal}}\ge 0$ is a regularization parameter, which
weighs the importance of $\mathcal{L}_{\mathsf{dal}}$ relative to $\mathcal{L}_{\mathsf{dec}}$.

\subsection{Conversion Algorithm}
\label{subsec:conversion}
At test time, a source speech feature sequence $\Vec{X}$ can be 
converted via \refalgo{default}.
\begin{algorithm}[t]
\caption{Default conversion algorithm}
\label{algo:default}
\begin{algorithmic}
\STATE $\Vec{Z}\leftarrow\Vec{X}$, ${\Vec{Y}} \leftarrow \Vec{0}$
\FOR{$l=1$ to $L$}
\STATE $\Vec{Z} \leftarrow \mathsf{Enc}_l(\Vec{Z})$
\ENDFOR
\FOR{$m=1$ to $M$}
\FOR{$l=1$ to $L$}
\STATE $\Vec{Y} \leftarrow \mathsf{Dec}_l(\Vec{Y},\Vec{Z})$
\ENDFOR
\STATE $\Vec{Y} \leftarrow [\Vec{0},\Vec{Y}]$
\ENDFOR
\STATE \textbf{return} $\Vec{Y}$
\end{algorithmic}
\end{algorithm}
Note that
in our model, the all-zero vector corresponds to the start-of-sequence token. 
As for the end-of-sequence token, we intentionally did not include it in the source and target feature sequences.
This is because we are assuming a situation where source speech features are constantly coming in and the conversion is performed online. 
In the following experiments, we set $M$ to a sufficiently large number (specifically, twice the length   $N$ of the source feature sequence) and regarded the time $m$ at which the attended time point (i.e., the peak of the attention distribution)
first reached $N$ as the end-of-utterance. 

Once $\Vec{Y}$ has been obtained, 
we adjust 
the mean and variance of the generated feature sequence 
so that they match the pretrained 
mean and variance of the feature vectors of the target speaker.
We can then generate a time-domain signal using the WORLD vocoder or any 
recently developed neural vocoder.

\begin{figure}[t!]
\centering
\begin{minipage}{.75\linewidth}
\centerline{\includegraphics[width=.99\linewidth]{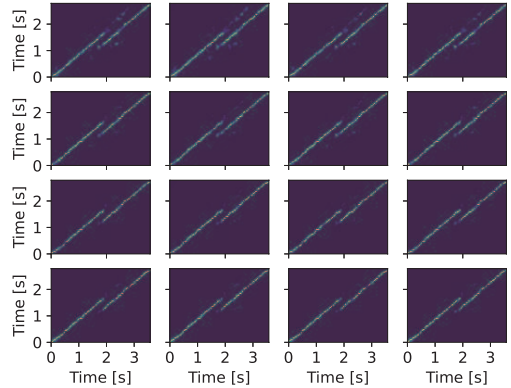}}
\end{minipage}
\caption{
Example of TSA matrices predicted 
using the original VTN with $L=4$ and $H=4$
trained from scratch (without pretraining).
Graph of column $h$ and row $l$ 
shows plot of $\Vec{A}_{l,h}$.
}
\label{fig:attention_pairwise_raw}
\end{figure}

However, as \reffig{attention_pairwise_raw} shows,
it transpired that with the model trained from scratch,
the attended time point 
did not always move forward monotonically and continuously 
at test time 
and occasionally 
made a sudden 
jump to a distant time point,
resulting in some segments being skipped or repeated, 
even though the DAL was considered in training.
In our previous work \cite{Huang2020Interspeech_VTN},
we proposed to introduce 
pretraining techniques exploiting auxiliary text labels 
to improve the behavior and performance of 
the conversion algorithm, as mentioned earlier.
In the next section, we present several ideas that can 
greatly improve the behavior of the VTN 
even without pretraining using text labels.

\section{Many-to-Many VTN}
\label{sec:improvedVTN}

\subsection{Many-to-Many Extension}
\label{subsec:multidomain}

\begin{figure}[t!]
\centering
\begin{minipage}{.7\linewidth}
  \centerline{\includegraphics[height=8.208cm]{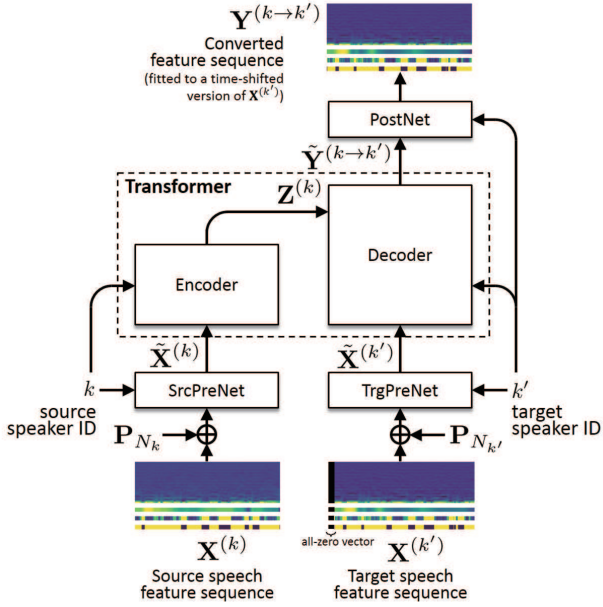}}
  \vspace{-1ex}
  \caption{Structure of many-to-many VTN.}
  \label{fig:multidomainVTN}
\end{minipage}
\vspace{-1ex}
\end{figure}

The first and main idea is a many-to-many extension of the VTN, 
which uses a single model to enable mappings among multiple speakers
by allowing the prenets, postnet,
encoder, and decoder to take source and target
speaker indices as additional inputs. 
The overall structure of the many-to-many VTN is shown in \reffig{multidomainVTN}.

Let $\Vec{X}^{(1)},\ldots,\Vec{X}^{(K)}$ 
be examples of the acoustic feature sequences of different speakers reading the same sentence.
Given a single pair of parallel utterances $\Vec{X}^{(k)}$ and $\Vec{X}^{(k')}$, 
where $k$ and $k'$ denote the source and target speaker indices (integers),
the source and target prenets take tuples $(\Vec{X}^{(k)},k)$ and $(\Vec{X}^{(k')},k')$ as the inputs
and produce modified feature sequences 
$\tilde{\Vec{X}}{}^{(k)}$ and $\tilde{\Vec{X}}{}^{(k')}$, respectively.
The encoder takes a tuple $(\tilde{\Vec{X}}{}^{(k)},k)$ as the input 
and produces a context vector sequence $\Vec{Z}^{(k)}$.
The decoder 
takes $(\tilde{\Vec{X}}{}^{(k')},\Vec{Z}^{(k)},k')$ 
as the input and produces a converted feature sequence $\tilde{\Vec{Y}}{}^{(k\rightarrow k')}$.
The postnet takes $(\tilde{\Vec{Y}}{}^{(k\rightarrow k')}, k')$ as the input 
and finally produces a modified version 
$\Vec{Y}{}^{(k\rightarrow k')}$ of $\tilde{\Vec{Y}}{}^{(k\rightarrow k')}$.
Each of the networks
incorporates the speaker index into its process
by modifying the input sequence, say $\Vec{X}$, via
\begin{align}
\Vec{S} &= \mathsf{repeat}(\mathsf{embed}(k)),
\label{eq:auxlayer1}
\\
\Vec{X} & \leftarrow [\Vec{X};\Vec{S}],
\label{eq:auxlayer2}
\end{align}
every time before feeding $\Vec{X}$ into the SA, FFN, or TSA sub-layers,
where
$\mathsf{embed}$ denotes an operation that retrieves a continuous vector from an integer input
and  
$\mathsf{repeat}$ denotes an operation that produces a vector sequence from an input vector
by simply repeating it along the time axis.
Again, we append an all-zero vector to the left end of $\Vec{X}{}^{(k')}$ 
\begin{align}
\Vec{X}{}^{(k')}\leftarrow [\Vec{0}, \Vec{X}{}^{(k')}].
\end{align}

The loss functions to be minimized
given this single training example are given as
\begin{align}
\mathcal{L}_{\mathsf{main}}^{(k,k')} &=
{\textstyle 
\frac{1}{N_{k'}}
}
\| [\Vec{Y}^{(k\rightarrow k')}]_{:,1:N_{k'}} - [\Vec{X}^{(k')}]_{:,2:N_{k'}+1} \|_1,
\label{eq:decloss_multidomain}
\\
\mathcal{L}_{\mathsf{dal}}^{(k,k')} &= 
{\textstyle \frac{1}{N_{k}N_{k'}HL}
\sum_h
\sum_l
}
\|\Vec{G}_{N_{k}\times N_{k'}}
\odot \Vec{A}_{l,h}^{(k,k')}\|_1,
\label{eq:recloss_multidomain}
\end{align}
where
$\Vec{A}_{l,h}^{(k,k')}$ denotes 
the TSA matrix of the $h$-th head in the TSA sub-layer 
in the $l$-th decoder layer.

With the above model,
we can also consider the case in which $k=k'$.
Minimizing the sum of the above losses under $k=k'$ 
encourages the model to let
the input feature sequence $\Vec{X}^{(k)}$ 
remain unchanged when $k$ and $k'$ indicate the same speaker.
We call this loss the identity mapping loss (IML).
The task of reconstructing an input sequence is relatively easier than that of finding the mapping from an input sequence to a different sequence. 
Using the IML as the training objective allows the model to concentrate on learning only the autoregressive behaviour of the decoder. 
This can contribute to making the model training easier. 
The total training loss 
including the IML thus becomes
\begin{align}
\mathcal{L} &=
\sum_{k,k'\neq k}
\mathbb{E}_{\Vec{X}^{(k)}\!,\Vec{X}^{(k')}}
\!
\big\{
\mathcal{L}_{\mathsf{all}}^{(k,k')}
\big\}
+
\lambda_{\mathsf{iml}}
\sum_{k}
\mathbb{E}_{\Vec{X}^{(k)}}
\!
\big\{
\mathcal{L}_{\mathsf{all}}^{(k,k)}
\big\},\nonumber\\
&{\rm where}~~
\mathcal{L}_{\mathsf{all}}^{(k,k')}=
\mathcal{L}_{\mathsf{main}}^{(k,k')}
+
\lambda_{\mathsf{dal}}
\mathcal{L}_{\mathsf{dal}}^{(k,k')},
\end{align}
where
$\mathbb{E}_{\Vec{X}^{(k)}\!,\Vec{X}^{(k')}}\{\cdot\}$
and $\mathbb{E}_{\Vec{X}^{(k)}}\{\cdot\}$
denote the sample means over all the training examples of parallel utterances of speakers $k$ and $k'$,
and $\lambda_{\mathsf{iml}}\ge 0$ is a regularization parameter, which weighs the importance of 
the IML. 
The significant effect of the IML is shown later.

\begin{figure}[t!]
\centering
\begin{minipage}{.75\linewidth}
\centerline{\includegraphics[width=.99\linewidth]{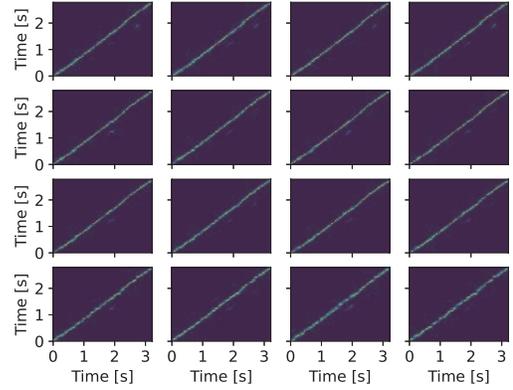}}
\end{minipage}
\caption{
Example of 
attention matrices predicted 
using many-to-many VTN with $L=4$ and $H=4$
trained from scratch.
Graph of column $h$ and row $l$ 
shows plot of $\Vec{A}_{l,h}$.
}
\label{fig:attention_multidomain_raw}
\end{figure}

\reffig{attention_multidomain_raw} 
shows examples of the TSA matrices
predicted using the many-to-many VTN from the same
test samples used in \reffig{attention_pairwise_raw}.
As these examples show, 
the predicted attention matrices obtained with the many-to-many VTN
exhibit more monotonic and continuous trajectories than those with the original VTN, 
demonstrating the impact of the many-to-many extension.

\subsection{Attention Windowing}

\begin{figure}[t!]
\centering
\begin{minipage}{.75\linewidth}
\centerline{\includegraphics[width=.99\linewidth]{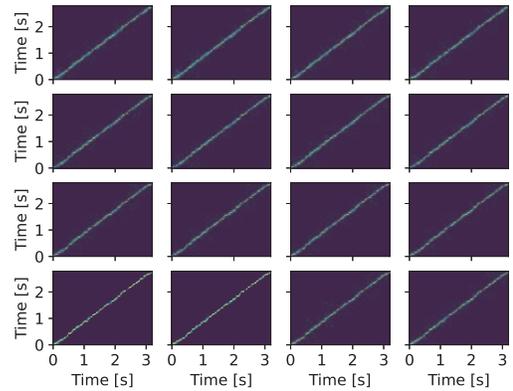}}
\end{minipage}
\caption{
Example of  
attention matrices predicted 
using the windowing technique
based on original VTN with $L=4$ and $H=4$
trained from scratch.
Graph of column $h$ and row $l$ 
shows plot of $\Vec{A}_{l,h}$.
}
\label{fig:attention_pairwise_forward}
\end{figure}

We present another idea 
that can be used alone or combined with the many-to-many extension
to improve the original VTN.
To assist the attended point to move forward monotonically and continuously at test time, 
we propose to modify \refalgo{default} by adopting an idea inspired by the technique called {\it windowing} \cite{Chorowski2015NIPS}.
Specifically, we limit the paths through which the attended point is allowed to move by forcing 
the attentions to all the time points
distant from the previous attended time point
to zeros. We assume
the attended time point to be
the peak of the attention distribution, given as 
the mean of 
all the TSA attention matrices in the TSA sub-layers in the decoder. 
This can be implemented by replacing 
\refeq{mhtsa_6} 
in the TSA sub-layer
in each decoder layer $l$
at the $m'$-th iteration of
the for-loop for $m=1,\ldots,M$ in 
the conversion process
with
\begin{align}
\hat{\Vec{A}}_{l,h} &= \mathsf{softmax}
\big(\textstyle{\frac{\Vec{K}_h^{\mathsf T}\Vec{Q}_h}{\sqrt{d}}} + \Vec{F} \big)
~(h=1,\ldots,H),
\end{align}
where the $(n,m)$-th element $f_{n,m}$ of $\Vec{F}$ is given by
\begin{align} 
f_{n,m} = 
\begin{cases}
-\infty&(m=m',~n=1,\ldots,\hat{n}-N_0)\\
-\infty&(m=m',~n=\hat{n}+N_1,\ldots,N)\\
0 &({\rm otherwise})
\end{cases},
\end{align}
so that all the elements of the last column of
the resulting $\hat{\Vec{A}}_{l,h}$ 
become zero except for the elements from row $\max(1,\hat{n}-N_0)$ to row $\min(\hat{n}+N_1,N)$.
$\Vec{Z}$ denotes the final output of the encoder,
$\Vec{X}$ and $\Vec{Y}$ denote the outputs of the previous and current sub-layers in the $l$-th decoder layer,
and $\hat{n}$ denotes 
the maximum point of the attention distribution
obtained at the $(m'-1)$-th iteration:
\begin{align}
\hat{n} = 
\begin{cases}
1&(m'=1)\\
\Argmax_{n} \frac{1}{LH} \sum_{l}\sum_{h}[\hat{\Vec{A}}_{l,h}]_{:,m'-1}&(m'\neq 1) 
\end{cases}.
\end{align}
Note that we set $N_0$ and $N_1$ at the nearest integers that 
correspond to 
$160$ and $320$ ms, respectively.
\reffig{attention_pairwise_forward} shows examples of 
the TSA matrices 
obtained with this algorithm from the same test samples
used in \reffig{attention_pairwise_raw}.
As these examples show, 
this algorithm 
was found to have a certain effect on generating 
reasonably monotonic and continuous attention trajectories
even without any modifications to the model structure of the original VTN. 

It should be noted that
we can also use the above algorithm as well as
the algorithm presented in \refsubsec{conversion}
for the many-to-many VTN, simply
by replacing $\mathsf{Enc}_l(\Vec{Z})$ and $\mathsf{Dec}_l(\Vec{Y},\Vec{Z})$
with
$\mathsf{Enc}_l(\Vec{Z},k)$ and $\mathsf{Dec}_l(\Vec{Y},\Vec{Z},k')$.

\medskip
\medskip
\subsection{Any-to-Many Conversion}
\label{subsec:many2one}

With both the one-to-one and many-to-many VTNs,
the source speaker must be known and specified at both training and test time.
However, in some cases 
we would need to 
handle {\it any}-to-many VC tasks, namely, to
convert the voice of an arbitrary speaker or 
an arbitrary speaking style that is not included in the training dataset.
Any-to-many VC is attractive in that it allows for input speech of unknown speakers without the need for extra processing, such as model retraining and adaptation.
It may also be useful as a voice normalization preprocessor for speaker-independent ASR.
Another important advantage of 
the many-to-many extension 
presented above 
is that it can be modified to handle any-to-many VC tasks
by intentionally not allowing the source prenet and encoder 
to take the source speaker index $k$ as inputs.
Namely, with this modified version,
the output sequence of each layer in these networks 
is directly passed to the next layer without going through
\refeqs{auxlayer1}{auxlayer2}.
We show later how well
this modified version performs on an any-to-many VC task
in which the source speaker is unseen in the training dataset.

\subsection{Real-Time System Settings}
\label{subsec:real-time}
It is important to be aware of real-time requirements when building VC systems. 
To let the VTN work in real-time, 
we need to make two modifications.
First,  
the source prenet and encoder must not use future information, as with 
the target prenet, decoder, and postnet during training. 
This can be implemented by 
constraining the convolution layers in the source prenet to be causal
and replacing \refeq{mhsa_5} with \refeq{mhsa_5_causal} 
for all the sub-layers in the encoder.
Second, since
the speaking rate and rhythm of input speech cannot be changed at test time, 
all the TSA matrices are simply set to identity matrices
so that the speaking rate and rhythm will be kept unchanged. 

\section{Experiments}
\label{sec:experiments}

\subsection{Experimental Settings}
\label{subsec:expcond}

To confirm the effectiveness of the ideas proposed in 
\refsec{improvedVTN}, 
we conducted objective and subjective evaluation experiments. 
We used the CMU Arctic database \cite{Kominek2004short},
which consists of recordings of four speakers, 
clb (female), 
bdl (male),
slt (female), and 
rms (male),
reading the same
1,132 phonetically balanced English utterances. 
We used all these speakers for training and evaluation.
Hence,  
there were a total of twelve different combinations of source and target speakers.
For each speaker, 
the first 1,000 and last 32 sentences of the 1,132 sentences
were used for training and evaluation, respectively. 
All the speech signals were sampled at 16 kHz. 
As already detailed in \refsubsec{feature},
for each utterance, 
the spectral envelope,
log $F_0$, 
coded aperiodicity,
and voiced/unvoiced information
were extracted every 8 ms 
using the
WORLD analyzer \cite{Morise2016short}. 
28 mel-cepstral coefficients (MCCs) were then extracted from 
each spectral envelope using the Speech Processing Toolkit (SPTK)\footnote{https://github.com/r9y9/pysptk}. 
The reduction factor $r$ was set to $3$.
Thus, $D=(28+3)\times 3=93$.

\vspace{-1ex}
\subsection{Network Architecture Details}

Dropouts with rate 0.1 were applied to the input sequences before
being fed into the source and target prenets and the postnet only at training time.
For the nonlinear activation function $\phi$ in each FFN sub-layer, 
we chose to use the GLU function since it yielded slightly better performance than 
the ReLU function.
The two prenets and the postnet 
were each designed using 
three 1D dilated convolution layers with kernel size $5$, each followed by a GLU activation function,
where weight normalization \cite{Salimans2016short} was applied to each layer. 
The channel number $d$ was set at 256 for the one-to-one VTN and 512 for the many-to-many VTN, respectively.
The middle channel number $d'$ of each FFN sub-layer
was set at 512 for the one-to-one VTN and 1024 for the many-to-many VTN, respectively.

\vspace{-1ex}
\subsection{Hyperparameter Settings}

$\lambda_{\mathsf{dal}}$ and $\lambda_{\mathsf{iml}}$
were set at 2000 and 1, respectively.
$\nu$ was set at 0.3 for both the one-to-one and many-to-many VTNs.
The $L_1$ norm $\|\Vec{X}\|_1$ used in 
\refeqs{decloss_pairwise}{decloss_multidomain} 
were defined as a weighted norm
$
\|\Vec{X}\|_1 =
\sum_{n=1}^{N}
\frac{1}{r}
\sum_{j=1}^{r}
\sum_{i=1}^{31} 
\gamma_i
|x_{ij,n}|$,
where 
$x_{1j,n},\ldots,x_{28j,n}$,
$x_{29j,n}$,
$x_{30j,n}$ and $x_{31j,n}$
denote the entries 
of $\Vec{X}$ corresponding to 
the 28 MCCs, log $F_0$, coded aperiodicity
and voiced/unvoiced indicator 
at time $n$,
and the weights 
were set at
$\gamma_1=\cdots=\gamma_{28}=\frac{1}{28}$,
$\gamma_{29}=\frac{1}{10}$,
$\gamma_{30}=\gamma_{31}=\frac{1}{50}$, respectively. 
$L$ and $H$ were set to 4 and 4 for the many-to-many version, and 
6 and 1 for the one-to-one version, respectively.
All the networks were trained simultaneously with random initialization.
Adam optimization \cite{Kingma2015short} was used for model training where 
the mini-batch size was 16 and 30,000 iterations were run. 
We configured each mini-batch to consist of only utterances of the same source-target speaker pair.
The learning rate and exponential decay rate
for the first moment 
for Adam were set to $1.0\times 10^{-4}$ and 0.9
for the many-to-many version with the PreLN architecture
and to $5.0\times 10^{-5}$ and 0.9 otherwise.
All these hyperparameters were tuned on a subset of 
the ATR Japanese speech database \cite{Kurematsu1990}, 
which consisted of 
503 phonetically balanced sentences
uttered by two male and two female speakers. 

\subsection{Objective Performance Measures}

The test dataset consisted 
of speech samples of each speaker reading the same sentences.
Thus, 
the quality of a converted feature sequence could be assessed by
comparing it with the feature sequence of the reference utterance.

\subsubsection{Mel-Cepstral Distortion}
Given two mel-cepstra,
we can use the mel-cepstral distortion (MCD)
to measure their difference.
We used the average of the MCDs 
taken along the DTW 
path between converted and reference feature sequences
as the objective performance measure for each test utterance.
Note that a smaller MCD indicates better performance.

\subsubsection{Log $F_0$ Correlation Coefficient}
To evaluate the log $F_0$ contour of converted speech, 
we used the correlation coefficient between the predicted and target log $F_0$ contours \cite{Hermes1998short}
as the objective performance measure. 
In the experiment, 
we used the average of the correlation coefficients taken over all the test utterances 
as the objective performance measure for log $F_0$ prediction.
Thus, the closer it is to 1, the better the performance.
We call this measure the log $F_0$ correlation coefficient (LFC).

\subsubsection{Local Duration Ratio}
To evaluate 
the speaking rate and rhythm of converted speech, 
we used 
the measure called the local duration ratio (LDR) \cite{Kameoka2018IEEE-TASLP_ConvS2S-VC}.
LDRs are defined as 
the local slopes of the DTW path between 
converted and reference utterances.
In the following, we use 
the mean absolute difference between the LDRs and 1 (in percentage)
as the overall measure for the LDRs.  
Thus, the closer it is to zero, the better the performance.
For example,
if the converted speech is 2 times faster than the reference speech,
the LDR will be 0.5 everywhere, and so its mean absolute difference from 1 will be $50\%$.

\vspace{-1ex}
\subsection{Baseline Methods}

\subsubsection{sprocket}
We chose the open-source VC method 
called sprocket \cite{Kobayashi2018short} for comparison in our experiments. 
To run this method,
we used the source code provided by the author\footnote{https://github.com/k2kobayashi/sprocket}.
Note that this method was used as a baseline in the
Voice Conversion Challenge (VCC) 2018 \cite{Lorenzo-Trueba2018short}.

\subsubsection{RNN-S2S-VC and ConvS2S-VC}

To compare different types of network architectures,
we also tested the RNN-based S2S model \cite{Tanaka2019ICASSP_AttS2S-VC}, 
inspired by the architecture introduced 
in an S2S model-based TTS system called Tacotron \cite{Wang2017short}, 
and the CNN-based model we presented previously 
\cite{Kameoka2018IEEE-TASLP_ConvS2S-VC}. 
We refer to these models as RNN-S2S-VC and ConvS2S-VC, respectively.

\noindent
{\bf RNN-S2S-VC:} 
Although the original Tacotron used mel-spectra as the acoustic features,
the baseline method was designed to use the same acoustic features as our method.
The architecture was specifically designed as follows.
The encoder consisted of a bottleneck fully connected prenet followed by a stack of 
$1\times 1$ 1D GLU convolutions and a bi-directional LSTM layer. 
The decoder was an autoregressive content-based attention network 
consisting of a  bottleneck fully connected prenet followed by 
a stateful LSTM  layer producing the attention query, which was then passed 
to a stack of 2 uni-directional residual LSTM layers, followed by a linear projection to generate the features. 

\begin{figure}[t!]
\centering
\begin{minipage}{.55\linewidth}
  \centerline{\includegraphics[height=7.5cm]{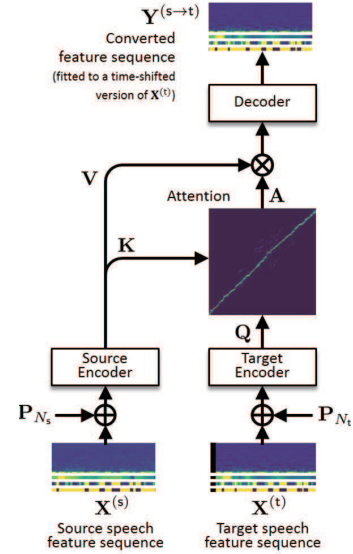}}
  \vspace{-1ex}
  \caption{Architecture of ConvS2S-VC}
  \label{fig:ConvS2S}
\end{minipage}
\vspace{-1ex}
\end{figure}

\noindent
{\bf ConvS2S-VC:} 
The ConvS2S model we implemented for this experiment 
consisted of source/target encoders and a decoder,
each of which had eight 1D GLU dilated convolution layers with kernel size of 5. 
We used single-step single-head scaled dot-product attention to 
compute attention distributions from the outputs
of the source/target encoders.
The convolutions in the target encoder and decoder 
were constrained to be causal, as with the target prenet and postnet 
in the one-to-one and many-to-many VTNs.
A residual connection and weight normalization were applied to each layer
in the three networks. 

We designed and implemented many-to-many extensions of 
the RNN-based and CNN-based models to compare them with the many-to-many VTN.

\subsection{Objective Evaluations}

\subsubsection{Ablation Studies}

\begin{table}[t]
\centering
\caption{Performance of 
one-to-one and many-to-many VTN
with PostLN and PreLN architectures
with and without WA.}
\begin{tabular}{l|l V{3} c V{3} c|c|c}
\thline

\multicolumn{2}{c V{3}}{\multirow{2}{*}{Versions}}&
\multirow{2}{*}{WA}&
\multicolumn{3}{c}{Measures}\\\cline{4-6}

\multicolumn{2}{c V{3}}{}& &MCD{\tiny (dB)}&LFC&LDR{\tiny (\%)}\\\thline

\multirow{4}{*}{\shortstack{one-to-\\one}}&\multirow{2}{*}{PostLN}&\multirow{1}{*}{--}&
$7.12$&$0.705$&$5.75$\\
 & &\multirow{1}{*}{\checkmark}&
$6.96$&$0.734$&$5.45$\\\cline{2-6}
 &\multirow{2}{*}{PreLN}&\multirow{1}{*}{--}&
$6.82$&$0.678$&$4.48$\\
 & &\multirow{1}{*}{\checkmark}&
$6.66$&$0.703$&$3.90$\\\hline

\multirow{4}{*}{\shortstack{many-to-\\many}}&\multirow{2}{*}{PostLN}&\multirow{1}{*}{--}&
$6.47$&$0.751$&$3.99$\\
 & &\multirow{1}{*}{\checkmark}&
$6.35$&$0.761$&$3.90$\\\cline{2-6}
 &\multirow{2}{*}{PreLN}&\multirow{1}{*}{--}&
$6.28$&$0.759$&$4.16$\\
 & &\multirow{1}{*}{\checkmark}&
$6.29$&$0.777$&$3.62$\\\thline
\end{tabular}
\label{tab:ablation1}
\end{table}
\begin{table}[t]
\centering
\caption{Performance of many-to-many VTN
trained with and without IML}
\begin{tabular}{l|l V{3} c V{3} c|c|c}
\thline
\multicolumn{2}{c V{3}}{\multirow{2}{*}{Versions}}&
\multirow{2}{*}{IML}&
\multicolumn{3}{c}{Measures}\\\cline{4-6}
\multicolumn{2}{c V{3}}{}& &MCD{\tiny (dB)}&LFC&LDR{\tiny (\%)}\\\thline

\multirow{4}{*}{\shortstack{many-to-\\many}}&
\multirow{2}{*}{PostLN}&
\multirow{1}{*}{--}&
$6.94$&$0.644$&$4.12$\\

& & \multirow{1}{*}{\checkmark}&
$6.35$&$0.761$&$3.90$\\\cline{2-6}

&\multirow{2}{*}{PreLN}&\multirow{1}{*}{--}&
$6.57$&$0.650$&$4.12$\\

& & \multirow{1}{*}{\checkmark}&
$6.28$&$0.792$&$2.51$\\\thline
\end{tabular}
\label{tab:ablation2}
\end{table}

\begin{table*}[t!]
\centering
\caption{MCDs (dB) obtained with baseline and proposed methods}
\begin{tabular}{l | l V{3} c|c|c|c|c|c|c}
\thline
\multicolumn{2}{c V{3}}{Speakers}&\multirow{2}{*}{sprocket}&\multicolumn{2}{c|}{RNN-S2S}&\multicolumn{2}{c|}{ConvS2S}&\multicolumn{2}{c}{VTN}\\\cline{1-2}\cline{4-9}
source&target&&one-to-one&many-to-many&one-to-one &many-to-many&one-to-one &many-to-many\\\thline
      &   bdl&$6.98$&$6.80$&$6.94$&$6.99$&$6.40$&$6.83$&$6.64$\\
   clb&   slt&$6.34$&$6.28$&$6.24$&$6.48$&$5.74$&$6.21$&$5.97$\\
      &   rms&$6.84$&$6.41$&$6.33$&$6.47$&$5.88$&$6.49$&$6.23$\\\hline
      &   clb&$6.44$&$6.33$&$6.20$&$6.61$&$5.79$&$6.17$&$6.03$\\
   bdl&   slt&$6.46$&$6.49$&$6.24$&$6.68$&$5.92$&$6.45$&$6.13$\\
      &   rms&$7.24$&$6.53$&$6.28$&$6.76$&$6.09$&$6.80$&$6.34$\\\hline
      &   clb&$6.21$&$6.20$&$6.21$&$6.41$&$5.69$&$6.20$&$5.91$\\
   slt&   bdl&$6.80$&$7.06$&$7.18$&$7.16$&$6.33$&$7.20$&$6.77$\\
      &   rms&$6.87$&$6.40$&$6.44$&$6.76$&$5.97$&$6.73$&$6.26$\\\hline
      &   clb&$6.43$&$6.36$&$6.26$&$6.38$&$5.88$&$6.63$&$5.94$\\
   rms&   bdl&$7.40$&$7.07$&$7.13$&$7.40$&$6.56$&$7.51$&$6.74$\\
      &   slt&$6.76$&$6.47$&$6.29$&$6.71$&$6.01$&$6.75$&$6.21$\\\hline
\multicolumn{2}{c V{3}}{All pairs}
             &$6.73$&$6.50$&$6.39$&$6.74$&$6.03$&$6.66$&$6.29$\\\thline
\end{tabular}
\label{tab:mcd_baseline_comp}
\end{table*}

\begin{table*}[t!]
\centering
\caption{LFCs obtained with baseline and proposed methods}
\begin{tabular}{l | l V{3} c|c|c|c|c|c|c}
\thline
\multicolumn{2}{c V{3}}{Speakers}&\multirow{2}{*}{sprocket}&\multicolumn{2}{c|}{RNN-S2S}&\multicolumn{2}{c|}{ConvS2S}&\multicolumn{2}{c}{VTN}\\\cline{1-2}\cline{4-9}
source&target&&one-to-one&many-to-many&one-to-one &many-to-many&one-to-one &many-to-many\\\thline
      &   bdl&$0.643$&$0.822$&$0.851$&$0.792$&$0.848$&$0.791$&$0.790$\\
   clb&   slt&$0.790$&$0.811$&$0.837$&$0.846$&$0.891$&$0.787$&$0.835$\\
      &   rms&$0.556$&$0.749$&$0.796$&$0.719$&$0.808$&$0.643$&$0.747$\\\hline
      &   clb&$0.642$&$0.738$&$0.809$&$0.752$&$0.828$&$0.785$&$0.767$\\
   bdl&   slt&$0.632$&$0.768$&$0.837$&$0.808$&$0.871$&$0.703$&$0.784$\\
      &   rms&$0.467$&$0.716$&$0.716$&$0.732$&$0.801$&$0.595$&$0.759$\\\hline
      &   clb&$0.820$&$0.748$&$0.774$&$0.795$&$0.849$&$0.750$&$0.772$\\
   slt&   bdl&$0.663$&$0.766$&$0.813$&$0.729$&$0.835$&$0.704$&$0.775$\\
      &   rms&$0.611$&$0.713$&$0.786$&$0.712$&$0.747$&$0.660$&$0.730$\\\hline
      &   clb&$0.632$&$0.785$&$0.815$&$0.809$&$0.829$&$0.632$&$0.717$\\
   rms&   bdl&$0.648$&$0.816$&$0.833$&$0.799$&$0.814$&$0.689$&$0.810$\\
      &   slt&$0.674$&$0.777$&$0.805$&$0.811$&$0.804$&$0.715$&$0.748$\\\hline
\multicolumn{2}{c V{3}}{All pairs}
      &$0.653$&$0.775$&$0.808$&$0.785$&$0.826$&$0.703$&$0.777$\\\thline   
\end{tabular}
\label{tab:lfc_baseline_comp}
\end{table*}
\begin{table*}[t!]
\centering
\caption{LDR deviations (\%) obtained with baseline and proposed methods}
\begin{tabular}{l | l V{3} c|c|c|c|c|c|c}
\thline
\multicolumn{2}{c V{3}}{Speakers}&\multirow{2}{*}{sprocket}&\multicolumn{2}{c|}{RNN-S2S}&\multicolumn{2}{c|}{ConvS2S}&\multicolumn{2}{c}{VTN}\\\cline{1-2}\cline{4-9}
source&target&&one-to-one&many-to-many&one-to-one &many-to-many&one-to-one &many-to-many\\\thline
      &   bdl&$17.66$&$3.52$&$3.22$&$3.04$&$3.34$&$3.33$&$2.34$\\
   clb&   slt&$ 9.74$&$2.34$&$2.70$&$0.86$&$4.38$&$3.90$&$4.28$\\
      &   rms&$ 3.24$&$2.70$&$3.76$&$4.18$&$6.79$&$5.57$&$2.82$\\\hline
      &   clb&$16.65$&$3.05$&$3.53$&$4.55$&$2.46$&$2.66$&$4.27$\\
   bdl&   slt&$ 4.58$&$4.18$&$4.21$&$6.10$&$3.52$&$3.77$&$3.04$\\
      &   rms&$15.20$&$2.90$&$2.21$&$3.78$&$3.42$&$2.69$&$3.00$\\\hline
      &   clb&$ 9.25$&$2.89$&$3.41$&$4.23$&$2.32$&$3.74$&$3.29$\\
   slt&   bdl&$ 5.52$&$2.35$&$2.04$&$3.47$&$3.72$&$3.69$&$3.03$\\
      &   rms&$11.46$&$3.06$&$5.00$&$3.66$&$5.57$&$6.88$&$5.41$\\\hline
      &   clb&$ 2.84$&$4.50$&$4.17$&$3.91$&$1.54$&$4.28$&$3.35$\\
   rms&   bdl&$17.76$&$4.68$&$3.19$&$3.41$&$4.15$&$3.31$&$3.40$\\
      &   slt&$11.95$&$4.74$&$3.74$&$3.61$&$4.48$&$4.28$&$4.27$\\\hline
\multicolumn{2}{c V{3}}{All pairs}
             &$10.60$&$3.30$&$3.38$&$3.77$&$3.69$&$3.90$&$3.62$\\\thline  
\end{tabular}
\label{tab:ldr_baseline_comp}
\end{table*}

We 
conducted ablation studies to 
confirm the effectiveness of 
the many-to-many VTN, 
IML, and attention windowing (AW),
and compared the performances obtained
with the PostLN and PreLN architectures.
It should be noted that 
the models trained without the DAL were unsuccessful 
in producing recognizable speech, 
possibly due to the limited amount of training data.
For this reason, we omit the results obtained when $\lambda_{\mathsf{dal}}=0$.

\reftab{ablation1}
lists the average MCDs, LFCs, and LDRs 
over the test samples
obtained with the one-to-one and many-to-many VTNs
with the PostLN and PreLN architectures
with and without AW. 
Note that all the results for the many-to-many version were obtained with the models trained using the IML.
We observed that
the effect of the many-to-many VTN was noticeable. 
Comparisons between with and without AW 
revealed that 
while the AW
showed a certain
effect in improving the one-to-one version in terms of all the measures, 
it was found to be only slightly effective for the many-to-many version. 
This may imply that the prediction of attentions
by the many-to-many version was already so successful that 
no correction by AW 
was necessary.
As for the PostLN and PreLN architectures,
the latter performed consistently better than the former, 
especially for the one-to-one version. 

\reftab{ablation2} shows the average MCDs, LFCs and LDRs 
over the test samples
obtained with the many-to-many VTN 
trained with and without the IML.
Note that all the results listed here are obtained using AW.
As these results indicate, the IML had a significant effect
on performance improvements in terms of
the MCD and LFC measures.

\subsubsection{Comparisons with Baseline Methods}

Tables \ref{tab:mcd_baseline_comp}, \ref{tab:lfc_baseline_comp} and \ref{tab:ldr_baseline_comp} 
show the MCDs, LFCs and LDRs obtained with the proposed and baseline methods.
It should be noted that  
sprocket was designed to only adjust the mean and variance of the log $F_0$ contour of input speech and keep the rhythm unchanged.
Hence, the performance gains over sprocket in terms of the LFC and LDR measures show
how well the competing methods are able to predict the $F_0$ contours and rhythms of target speech.
As the results indicate, 
all the S2S models performed better than sprocket 
in terms of the LFC and LDR measures, thus demonstrating the ability to
properly convert the prosodic features in speech. 
They also performed better than or comparably to sprocket 
in terms of the MCD measure. 
It is worth noting that 
the many-to-many extension was found to be significantly effective 
for all the architecture types of the S2S models.
It is interesting to compare the performance of
the many-to-many versions of RNN-S2S-VC, ConvS2S-VC and VTN.
The many-to-many version of ConvS2S-VC performed best in terms of the 
MCD and LFC measures whereas 
the many-to-many RNN-S2S-VC performed best in terms of the 
LDR measure.
This may indicate that 
the strengths of S2S models 
can vary depending on the type of architecture.
 
As mentioned earlier, 
one important advantage of the transformer architecture 
over its RNN counterpart 
is that it can be trained efficiently 
thanks to its parallelizable structure.
In fact, while 
it took about 30 hours and 50 hours to train 
the one-to-one and many-to-many versions of the RNN-S2S-VC model, 
it only took about 3 hours and 5 hours to train 
these two versions of the VTN
under the current experimental settings.
We implemented all the methods in PyTorch
and used a single Tesla V100 GPU with a 32.0GB memory
for training each model.

\begin{table}[t]
\centering
\caption{Performance of many-to-many VTN
with any-to-many setting 
under open-set condition
and sprocket under closed-set condition
tested on same samples}
(a) any-to-many VTN\\
\begin{tabular}{l| l V{3} c|c|c}
\thline
\multicolumn{2}{c V{3}}{Speaker pair}&\multicolumn{3}{c}{Measures}\\\hline
source&target&MCD{\tiny (dB)}&LFC&LDR{\tiny (\%)}\\\thline
\multirow{4}{*}{lnh} 
   &clb&$6.49$&$0.690$&$2.18$\\
   &bdl&$7.24$&$0.636$&$4.44$\\
   &slt&$6.59$&$0.693$&$4.40$\\
   &rms&$6.87$&$0.466$&$8.65$\\\hline
\multicolumn{2}{c V{3}}{All pairs}&$6.71$&$0.630$&$4.41$\\\thline
\end{tabular}
\\
\medskip\medskip
(b) sprocket\\
\begin{tabular}{l| l V{3} c|c|c}
\thline
\multicolumn{2}{c V{3}}{Speaker pair}&\multicolumn{3}{c}{Measures}\\\hline
source&target&MCD{\tiny (dB)}&LFC&LDR{\tiny (\%)}\\\thline
\multirow{4}{*}{lnh} 
   &clb&$6.76$&$0.716$&$6.61$\\
   &bdl&$8.26$&$0.523$&$13.38$\\
   &slt&$6.62$&$0.771$&$5.72$\\
   &rms&$7.22$&$0.480$&$4.87$\\\hline
\multicolumn{2}{c V{3}}{All pairs}&$7.21$&$0.579$&$7.61$\\\thline
\end{tabular}
\label{tab:all_m2o_unseen}
\end{table}

\subsubsection{Performance of any-to-many VTN}

We evaluated the performance 
of the any-to-many VTN described in \refsubsec{many2one}
under an open-set condition 
where the speaker of the test utterances is unseen in the training data.
We used the utterances of speaker lnh (female) as the test input speech.
The results are listed in \reftab{all_m2o_unseen} (a).
For comparison, 
\reftab{all_m2o_unseen} (b) lists the results of 
sprocket performed on the same speaker pairs
under a speaker-dependent closed-set condition.
As these results indicate, the any-to-many VTN
still performed better than sprocket, even though sprocket had an advantage in
both the training and test conditions.

\begin{table}[t]
\caption{Performance of many-to-many VTN
with real-time system setting}
\centering
\begin{tabular}{l| l V{3} c|c}
\thline
\multicolumn{2}{c V{3}}{Speaker pair}&\multicolumn{2}{c}{Measures}\\\hline
source&target&MCD{\tiny (dB)}&LFC\\\thline
\multirow{3}{*}{clb} 
   &bdl&$7.27$&$0.735$\\
   &slt&$6.13$&$0.791$\\
   &rms&$6.75$&$0.693$\\\hline
\multirow{3}{*}{bdl} 
   &clb&$6.36$&$0.685$\\
   &slt&$6.61$&$0.715$\\
   &rms&$6.61$&$0.660$\\\hline
\multirow{3}{*}{slt} 
   &clb&$6.12$&$0.743$\\
   &bdl&$7.10$&$0.673$\\
   &rms&$6.55$&$0.609$\\\hline
\multirow{3}{*}{rms} 
   &clb&$6.06$&$0.737$\\
   &bdl&$7.22$&$0.612$\\
   &slt&$6.60$&$0.730$\\\hline
\multicolumn{2}{c V{3}}{All pairs}&$6.58$&$0.703$\\\thline
\end{tabular}
\label{tab:all_causal}
\end{table}

\subsubsection{Performance with Real-Time System Setting}

We evaluated the MCDs and LFCs obtained with
the many-to-many VTN with the real-time system setting described in \refsubsec{real-time}.
The results are shown in Table \ref{tab:all_causal}.
It is worth noting that
it performed only slightly worse than with the default setting
despite  
the restrictions related to the real-time system setting
and performed still better than sprocket 
in terms of the MCD and LFC measures.

\begin{table*}[t!]
\centering
\caption{MCDs (dB) obtained with one-to-one and many-to-many VTNs under different training data size conditions}
\begin{tabular}{l | l V{3} c|c|c|c|c|c}
\thline
\multicolumn{2}{c V{3}}{Speakers}&\multicolumn{2}{c|}{1000 training utterances}&\multicolumn{2}{c|}{500 training utterances}&\multicolumn{2}{c}{250 training utterances}\\\cline{1-2}\cline{3-8}
source&target&one-to-one&many-to-many&one-to-one&many-to-many&one-to-one&many-to-many\\\thline
      &   bdl&$6.83$&$6.64$&$7.04$&$7.03$&$7.49$&$7.55$\\
   clb&   slt&$6.21$&$5.97$&$6.44$&$6.27$&$6.79$&$6.79$\\
      &   rms&$6.49$&$6.23$&$6.72$&$6.56$&$7.22$&$7.02$\\\hline
      &   clb&$6.17$&$6.03$&$6.53$&$6.41$&$7.37$&$6.80$\\
   bdl&   slt&$6.45$&$6.13$&$6.72$&$6.48$&$7.19$&$6.97$\\
      &   rms&$6.80$&$6.34$&$7.24$&$6.71$&$7.67$&$7.40$\\\hline
      &   clb&$6.20$&$5.91$&$6.36$&$6.28$&$6.92$&$6.67$\\
   slt&   bdl&$7.20$&$6.77$&$7.21$&$7.24$&$7.76$&$7.82$\\
      &   rms&$6.73$&$6.26$&$6.94$&$6.73$&$7.67$&$7.24$\\\hline
      &   clb&$6.63$&$5.94$&$6.86$&$6.40$&$7.36$&$6.97$\\
   rms&   bdl&$7.51$&$6.74$&$7.39$&$7.13$&$8.11$&$7.83$\\
      &   slt&$6.75$&$6.21$&$7.23$&$6.45$&$7.65$&$7.04$\\\hline
\multicolumn{2}{c V{3}}{All pairs}
             &$6.66$&$6.29$&$6.89$&$6.64$&$7.47$&$7.06$\\\thline
\end{tabular}
\label{tab:mcd_training_datasize_comp}
\end{table*}

\begin{table*}[t!]
\centering
\caption{LFCs obtained with one-to-one and many-to-many VTNs under different training data size conditions}
\begin{tabular}{l | l V{3} c|c|c|c|c|c}
\thline
\multicolumn{2}{c V{3}}{Speakers}&\multicolumn{2}{c|}{1000 training utterances}&\multicolumn{2}{c|}{500 training utterances}&\multicolumn{2}{c}{250 training utterances}\\\cline{1-2}\cline{3-8}
source&target&one-to-one&many-to-many&one-to-one&many-to-many&one-to-one&many-to-many\\\thline
      &   bdl&$0.791$&$0.790$&$0.758$&$0.689$&$0.709$&$0.676$\\
   clb&   slt&$0.787$&$0.835$&$0.801$&$0.827$&$0.815$&$0.803$\\
      &   rms&$0.643$&$0.747$&$0.693$&$0.675$&$0.631$&$0.574$\\\hline
      &   clb&$0.785$&$0.767$&$0.797$&$0.725$&$0.721$&$0.719$\\
   bdl&   slt&$0.703$&$0.784$&$0.756$&$0.818$&$0.759$&$0.769$\\
      &   rms&$0.595$&$0.759$&$0.647$&$0.724$&$0.573$&$0.503$\\\hline
      &   clb&$0.750$&$0.772$&$0.804$&$0.771$&$0.746$&$0.720$\\
   slt&   bdl&$0.704$&$0.775$&$0.715$&$0.665$&$0.578$&$0.639$\\
      &   rms&$0.660$&$0.730$&$0.658$&$0.655$&$0.567$&$0.533$\\\hline
      &   clb&$0.632$&$0.717$&$0.682$&$0.704$&$0.591$&$0.673$\\
   rms&   bdl&$0.689$&$0.810$&$0.754$&$0.628$&$0.637$&$0.513$\\
      &   slt&$0.715$&$0.748$&$0.630$&$0.765$&$0.666$&$0.73$\\\hline
\multicolumn{2}{c V{3}}{All pairs}
             &$0.703$&$0.777$&$0.717$&$0.723$&$0.668$&$0.688$\\\thline
\end{tabular}
\label{tab:lfc_training_datasize_comp}
\end{table*}

\begin{table*}[t!]
\centering
\caption{LDR deviations (\%) obtained with one-to-one and many-to-many VTNs under different training data size conditions}
\begin{tabular}{l | l V{3} c|c|c|c|c|c}
\thline
\multicolumn{2}{c V{3}}{Speakers}&\multicolumn{2}{c|}{1000 training utterances}&\multicolumn{2}{c|}{500 training utterances}&\multicolumn{2}{c}{250 training utterances}\\\cline{1-2}\cline{3-8}
source&target&one-to-one&many-to-many&one-to-one&many-to-many&one-to-one&many-to-many\\\thline
      &   bdl&$3.33$&$2.34$&$12.10$&$8.47$&$9.03$&$12.30$\\
   clb&   slt&$3.90$&$4.28$&$4.87$&$2.66$&$4.71$&$4.16$\\
      &   rms&$5.57$&$2.82$&$3.22$&$3.73$&$4.24$&$7.66$\\\hline
      &   clb&$2.66$&$4.27$&$7.25$&$3.67$&$4.69$&$2.61$\\
   bdl&   slt&$3.77$&$3.04$&$8.19$&$3.34$&$8.29$&$3.06$\\
      &   rms&$2.69$&$3.00$&$9.44$&$5.61$&$8.12$&$1.80$\\\hline
      &   clb&$3.74$&$3.29$&$3.05$&$3.01$&$3.37$&$3.36$\\
   slt&   bdl&$3.69$&$3.03$&$10.15$&$8.49$&$11.51$&$9.73$\\
      &   rms&$6.88$&$5.41$&$5.18$&$2.26$&$7.59$&$3.44$\\\hline
      &   clb&$4.28$&$3.35$&$2.94$&$3.21$&$5.33$&$4.47$\\
   rms&   bdl&$3.31$&$3.40$&$10.10$&$10.65$&$11.99$&$13.88$\\
      &   slt&$4.28$&$4.27$&$6.94$&$3.41$&$5.57$&$3.82$\\\hline
\multicolumn{2}{c V{3}}{All pairs}
             &$3.90$&$3.62$&$6.35$&$4.13$&$6.70$&$4.63$\\\thline
\end{tabular}
\label{tab:ldr_training_datasize_comp}
\end{table*}

\subsubsection{Impact of training data size}

To evaluate the impact of the training data size,
we compared the performance of the one-to-one and many-to-many VTNs
trained using 1,000, 500, and  250 utterances for each speaker, respectively.
These results are shown in 
Tables \ref{tab:mcd_training_datasize_comp}, 
\ref{tab:lfc_training_datasize_comp}, and \ref{tab:ldr_training_datasize_comp}.
We can confirm that the performance of both versions degrades as expected in terms of all the measures as the training data size decreases. 
More importantly, we can see that the many-to-many VTN trained using only 500 utterances for each speaker performed comparably or slightly better than the one-to-one VTN and sprocket trained using 1,000 utterances for each speaker.
This implies the fact that in training the mapping between a certain speaker pair using $500\times 2$ utterances, $500\times 2$ utterances of the remaining two speakers are as valuable as another $500\times 2$ utterances of that speaker pair, if leveraged efficiently. 

\begin{figure*}[t!]
\centering
\begin{minipage}[t]{.32\linewidth}
\centering
\centerline{\includegraphics[width=.98\linewidth]{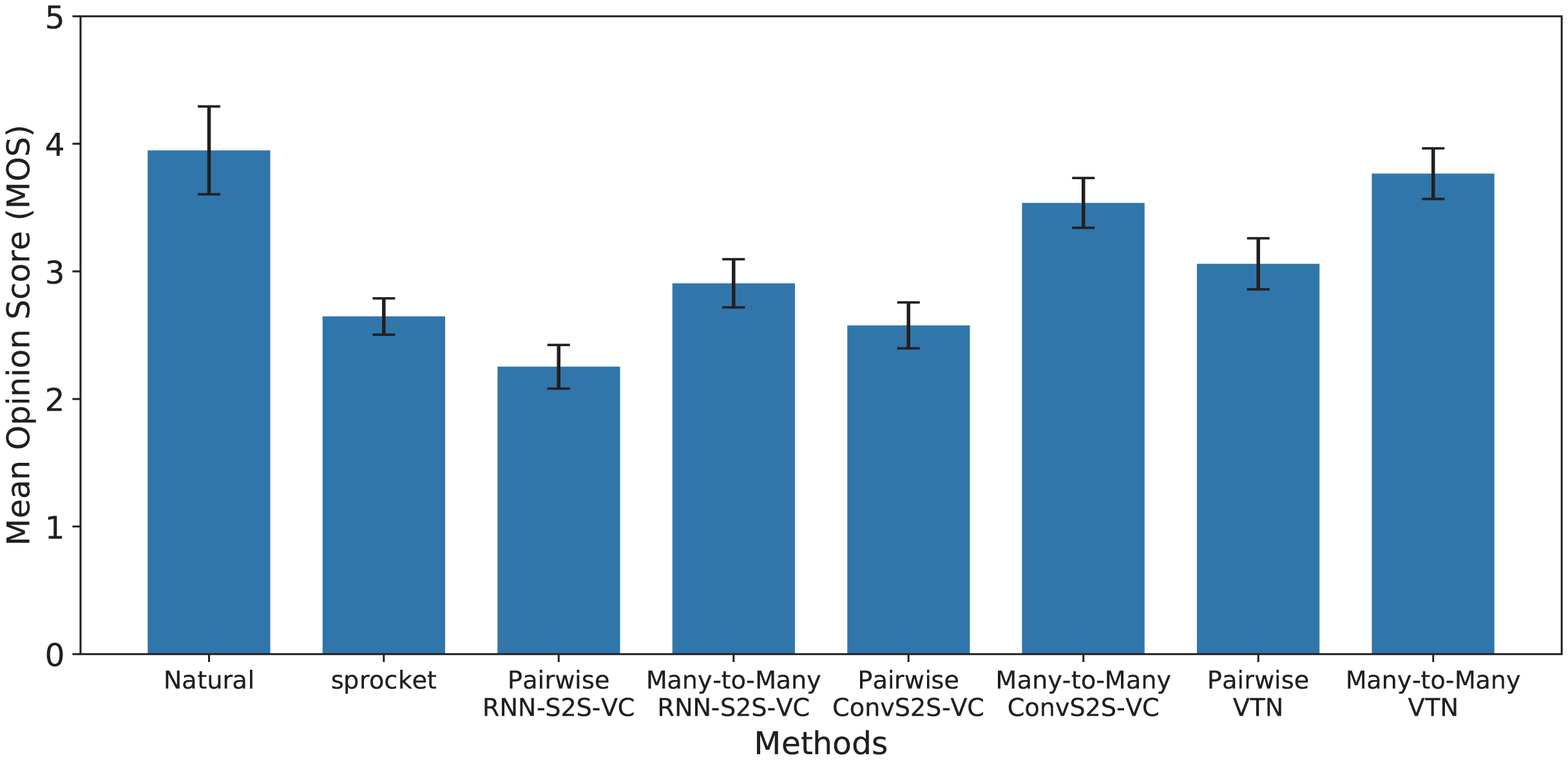}}
\end{minipage}
\centering
\begin{minipage}[t]{.32\linewidth}
\centering
\centerline{\includegraphics[width=.98\linewidth]{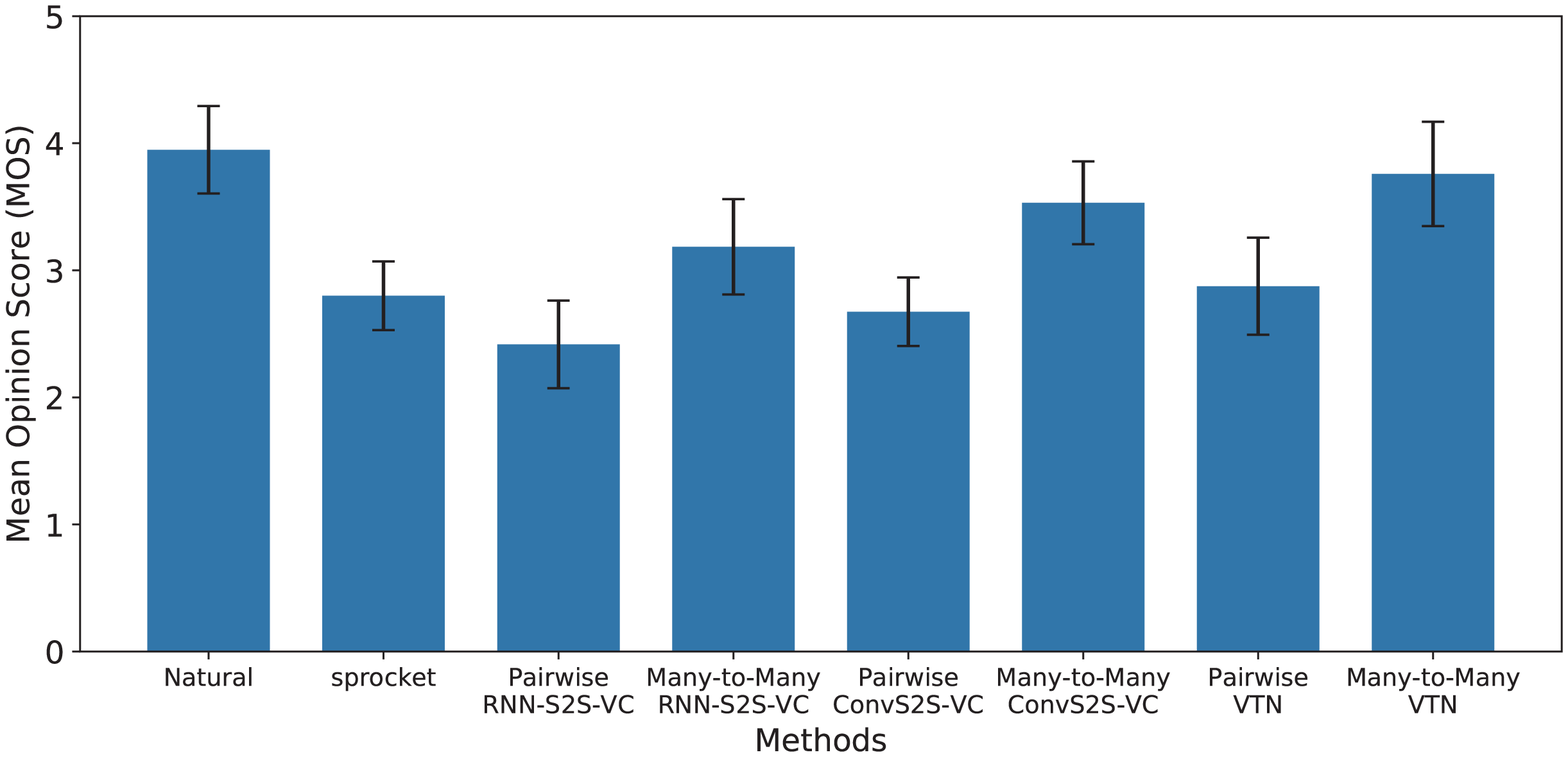}}
\end{minipage}
\centering
\begin{minipage}[t]{.32\linewidth}
\centering
\centerline{\includegraphics[width=.98\linewidth]{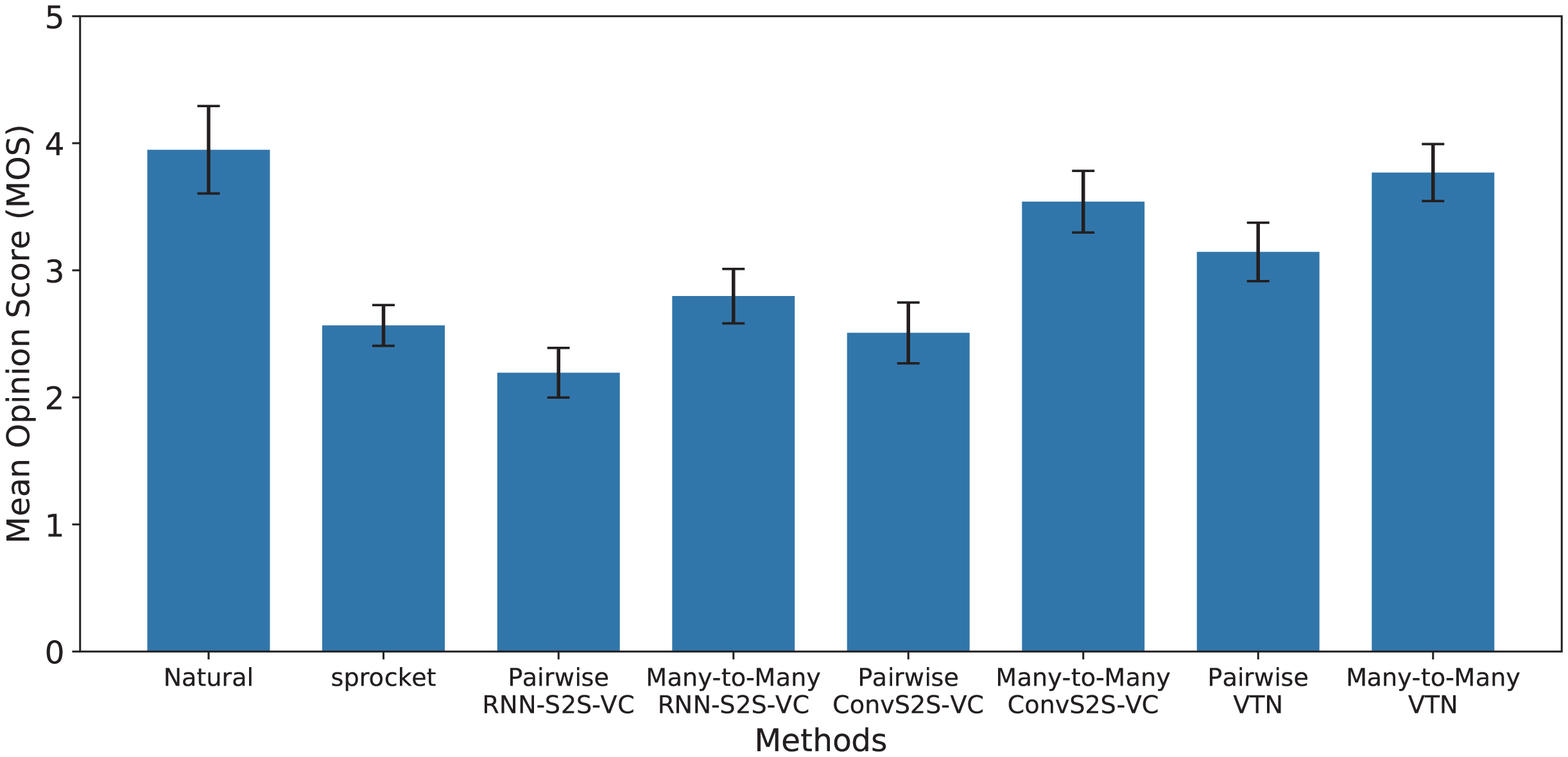}}
\end{minipage}
\\
\vspace{-1ex}
  \caption{Sound quality scores averaged across all speaker, intra-gender, and inter-gender pairs, respectively (from left to right).}
  \label{fig:MOS_qlt}
  \medskip\medskip
\centering
\begin{minipage}[t]{.32\linewidth}
\centering
\centerline{\includegraphics[width=.98\linewidth]{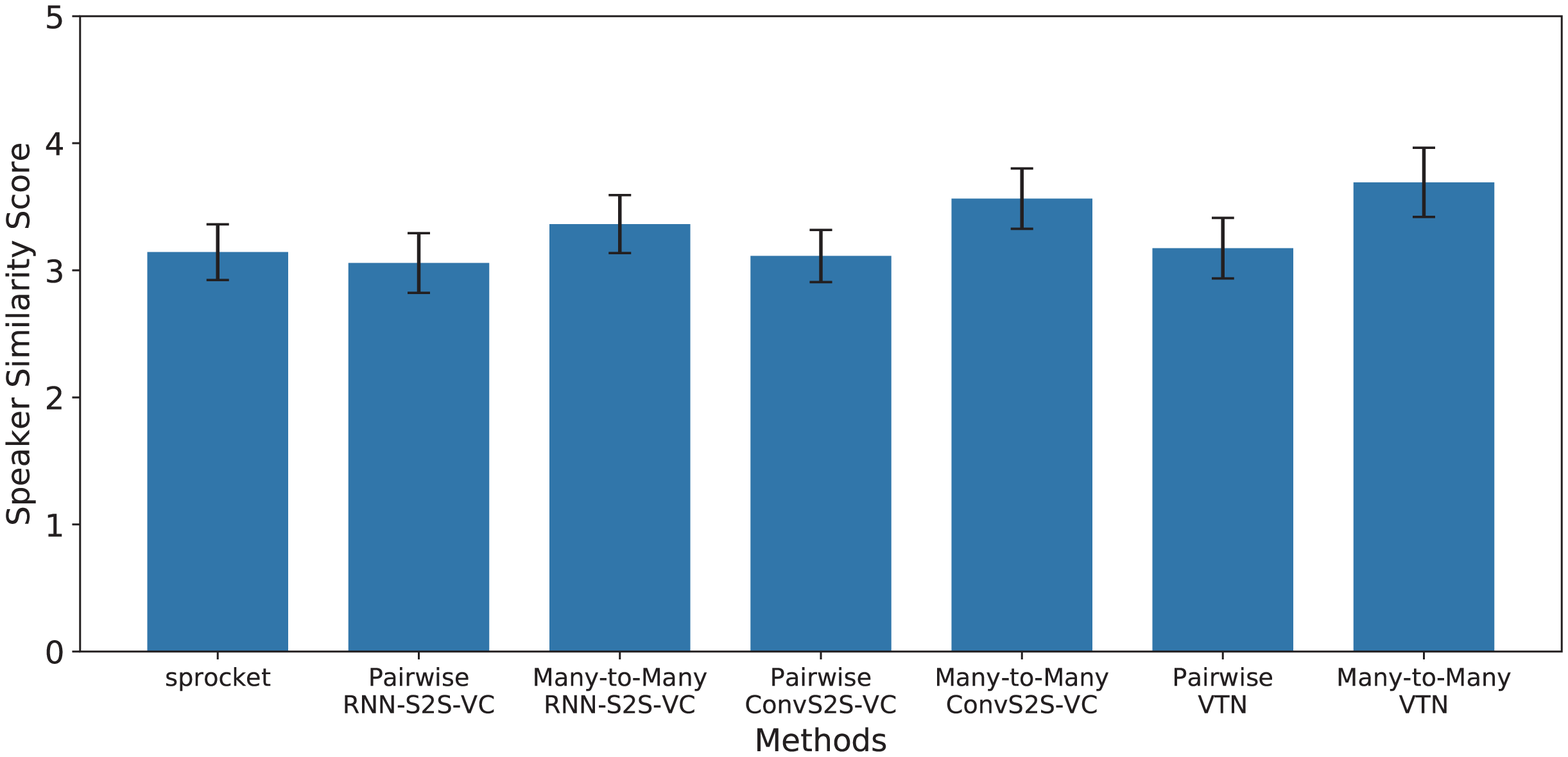}}
\end{minipage}
\centering
\begin{minipage}[t]{.32\linewidth}
\centering
\centerline{\includegraphics[width=.98\linewidth]{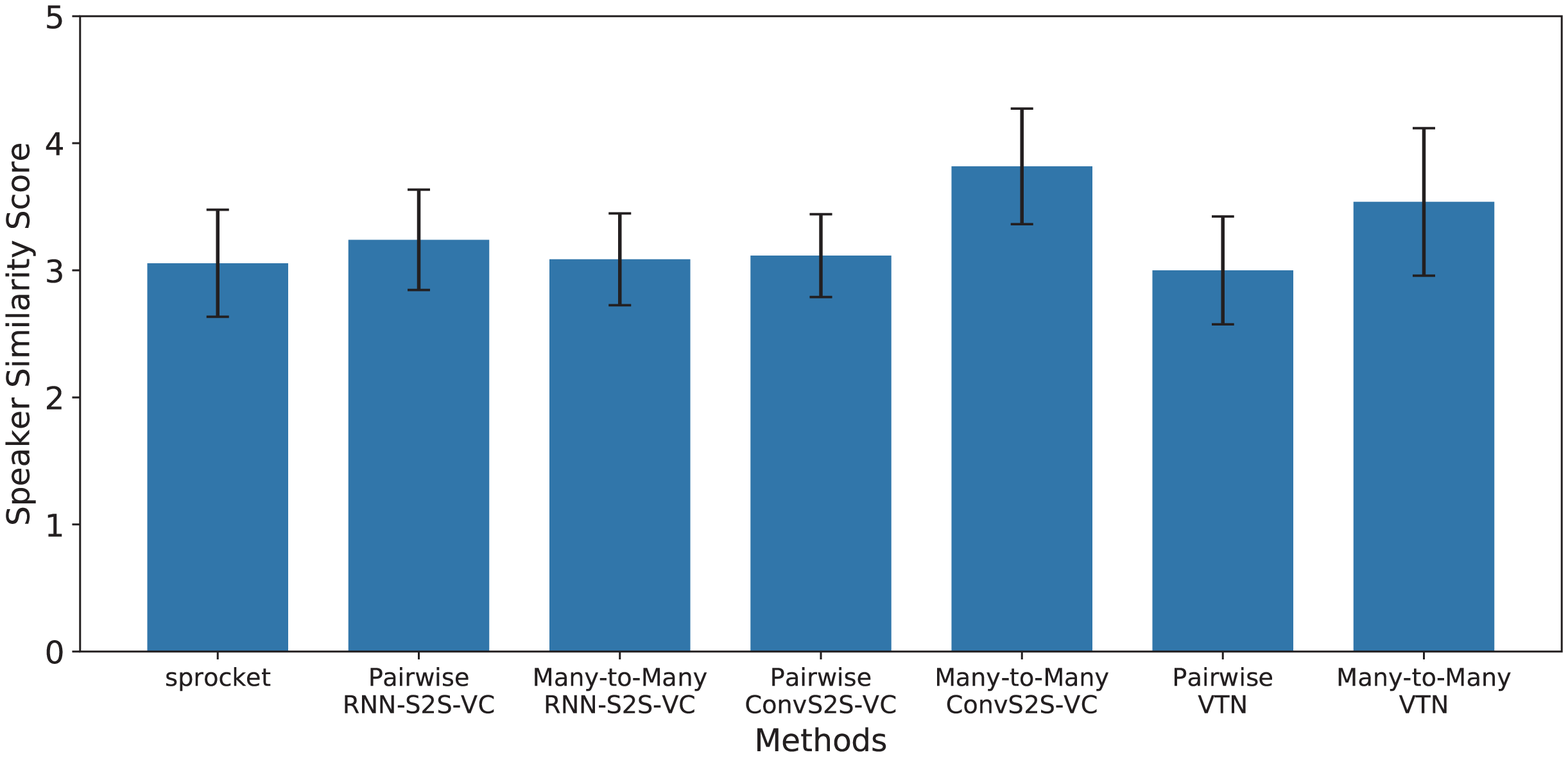}}
\end{minipage}
\centering
\begin{minipage}[t]{.32\linewidth}
\centering
\centerline{\includegraphics[width=.98\linewidth]{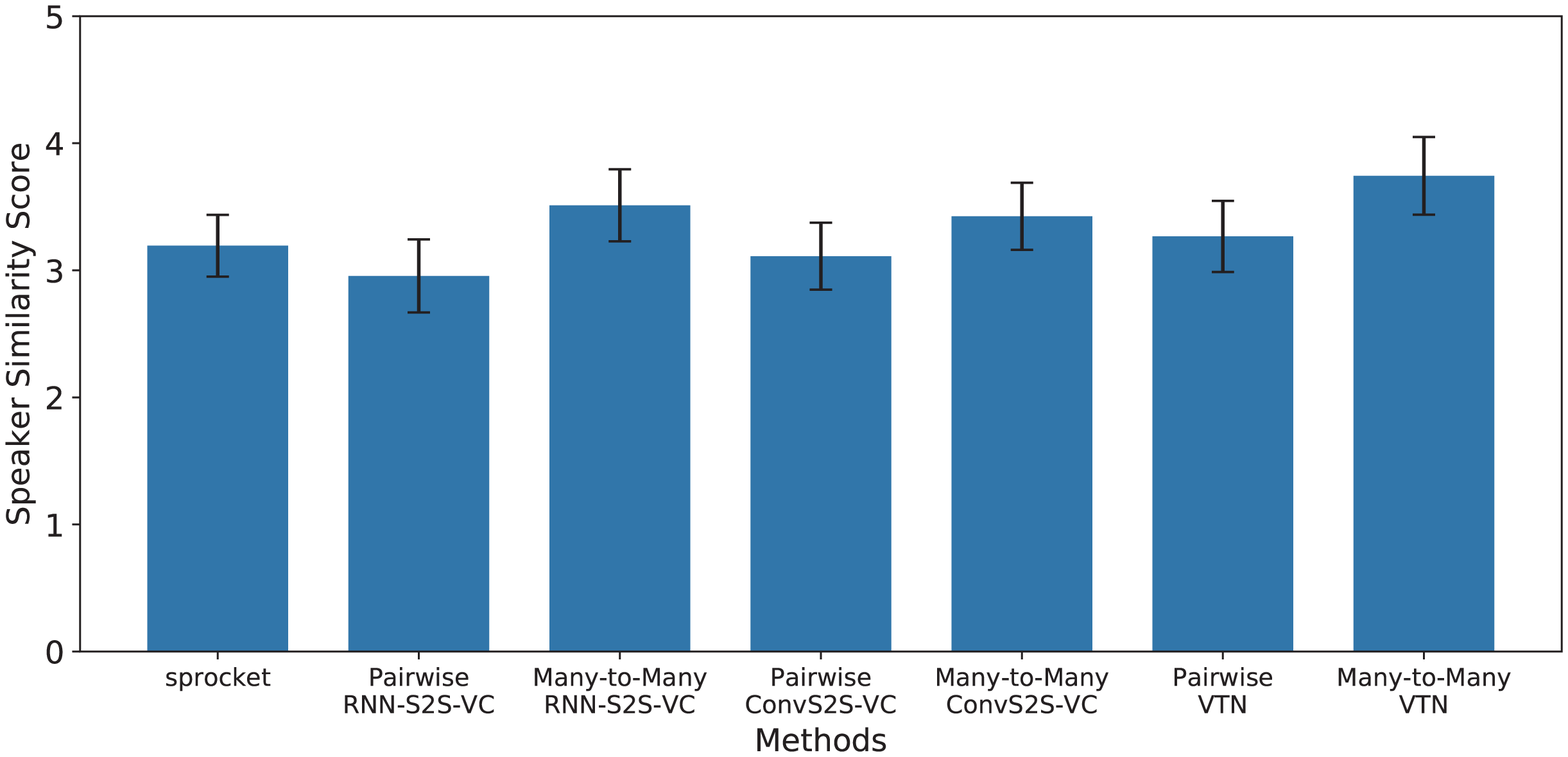}}
\end{minipage}
\vspace{-1ex}
\caption{Speaker similarity scores averaged across all speaker, intra-gender, and inter-gender pairs, respectively (from left to right).}
\label{fig:MOS_sim}
\end{figure*}

\subsection{Subjective Listening Tests}

We conducted subjective listening tests to compare the sound quality 
and speaker similarity 
of the converted speech samples
obtained with the proposed and baseline methods.
For these tests, we used 32 speech samples generated by each method for each source-target speaker pair.

With the sound quality test, we evaluated the mean opinion
score (MOS) for each speech sample. In this test, 
we included the speech samples 
synthesized in the same manner as the proposed and baseline methods (namely, the WORLD synthesizer) using
the acoustic features directly extracted from real speech samples. 
We also included speech samples produced using the one-to-one and many-to-many versions of 
RNN-S2S-VC, ConvS2S-VC, and VTN and sprocket in the stimuli.
Twenty listeners participated in our listening tests. 
Each listener was 
asked to evaluate the naturalness
by selecting 5: Excellent, 4: Good, 3: Fair, 2: Poor, or 1: Bad for each utterance.
The scores averaged across all, intra-gender, and inter-gender pairs are shown in \reffig{MOS_qlt}.
The one-to-one VTN performed better than 
sprocket and 
the one-to-one versions of the other S2S models. 
We also confirmed that 
the many-to-many extension had a significant effect in  
improving the audio quality of all the S2S models. 
It is worth noting that 
the many-to-many VTN performed better than 
all the competing methods including 
the many-to-many version of ConvS2S-VC,
even though the many-to-many version of ConvS2S-VC 
was found to outperform 
the many-to-many VTN 
in terms of the MCD and LFC measures through 
the objective evaluation experiments,
as reported earlier. 
According to the two-sided Mann-Whitney test performed on the MOS scores of the many-to-many VTN and each of the remaining methods, the $p$-values for all the pairs except for the many-to-many VTN and many-to-many ConvS2S-VC pair were less than 0.05, indicating that the many-to-many VTN performed significantly better than all the competing methods except the many-to-many ConvS2S-VC in terms of sound quality.

With the speaker similarity test, 
each listener was given a converted speech sample and 
a real speech sample of the corresponding target speaker 
and was asked to evaluate how likely they are to be produced by the same speaker by selecting 
5: Definitely, 4: Likely, 3: Fairly likely, 2: Not very likely, or 1: Unlikely.
We used converted speech samples generated by the one-to-one and many-to-many versions 
of RNN-S2S-VC and ConvS2S-VC, and sprocket for comparison as with the sound quality test. 
The scores averaged across all, intra-gender, and inter-gender pairs are shown in \reffig{MOS_sim}.
The many-to-many versions of ConvS2S-VC and VTN 
performed comparably and 
slightly better than all other methods.
According to the two-sided Mann-Whitney test, the many-to-many VTN was found to perform significantly better than the one-to-one VTN, one-to-one ConvS2S-VC, one-to-one RNN-S2S-VC, and sprocket in terms of speaker similarity.

Audio samples of the one-to-one and many-to-many VTNs 
are available on the web\footnote{http://www.kecl.ntt.co.jp/people/kameoka.hirokazu/Demos/vtn/index.html}. 

\section{Conclusion}

We proposed an extension of VTN, 
which provides 
the flexibility of handling  many-to-many, any-to-many, and real-time VC tasks 
without relying on ASR models and text annotations. 
Through ablation studies, we confirmed the effectiveness of each of
the proposed ideas.
Objective and subjective evaluation experiments on a speaker identity conversion task
showed that the proposed method could perform better than baseline methods. 

Although we used the WORLD vocoder for waveform generation in the above experiments, 
using a neural vocoder instead could significantly improve the quality of the converted speech.
Rather than simply performing feature mapping and then using a neural vocoder to generate waveforms, we believe that further improvements could be made by integrating the VTN and a neural vocoder into a single model so that the whole model can be trained end-to-end.

Zero-shot VC is a task of converting input speech to the voice or speaking style of an unseen speaker by looking at only a few of his/her utterances \cite{Qian2019}.
Although in the many-to-many VTN, the target voice or speaking style is specified via a target speaker embedding vector, the embedding vector currently used for target speaker conditioning is nongeneralizable to unseen speakers. 
However, as proposed in \cite{Qian2019}, replacing the embedding vector with one used for speaker verification \cite{Wan2018ICASSP} may allow our model to handle zero-shot VC.





\ifCLASSOPTIONcaptionsoff
  \newpage
\fi



\bibliographystyle{IEEEtran}
\bibliography{Kameoka2020IEEETrans_VTN}

\end{document}